\documentclass[12pt,english,article,amsmath,amssymb]{article}
\usepackage{geometry}
\usepackage{graphicx,graphics}
\usepackage{dcolumn}
\usepackage{amsmath,amssymb,amsfonts}
\usepackage{latexsym,verbatim}
\usepackage{bm,bbm}
\usepackage{color}
\usepackage{ulem}

\usepackage[T1]{fontenc}
\usepackage[latin1]{inputenc}
\usepackage{dcolumn}
\usepackage{color}
\usepackage{babel}

\newcommand{\bR}{{\bm R}}

\usepackage[percent]{overpic}

\newcommand{\br}{{\bm r}}

\definecolor{dgreen}{rgb}{0,0.5,0}
\begin{document}

\title{
1/f critical current noise in short ballistic graphene Josephson junctions
}
\author{Francesco M.D. Pellegrino$^{1,2}$\footnote{Corresponding author: francesco.pellegrino@ct.infn.it}, 
Giuseppe Falci$^{1,2,3}$ and Elisabetta Paladino$^{1,2,3}$}
\date{}
\maketitle
\noindent$^1$ Dipartimento di Fisica e Astronomia "Ettore Majorana", Universit\`{a} di Catania,\\ Via S.~Sofia 64, I-95123 Catania, Italy. \\
$^2$ INFN, Sez. Catania, I-95123 Catania, Italy. \\
$^3$ CNR-IMM, Via S. Sofia 64, I-95123 Catania, Italy.\\
\maketitle

\begin{abstract}
Short ballistic graphene Josephson junctions sustain superconducting current with a 
non-sinusoidal current-phase relation up to a critical current threshold.
The current-phase relation, arising from proximitized superconductivity, is gate-voltage tunable and exhibits 
peculiar skewness observed in high quality  graphene superconductors heterostructures with clean interfaces.
These properties make graphene Josephson junctions promising sensitive quantum probes of microscopic 
fluctuations underlying  transport in two-dimensions.  
We show that the power spectrum of the critical current fluctuations 
has a characteristic $1/f$ dependence on frequency, $f$, probing  two points and higher correlations of 
carrier density fluctuations of the graphene channel induced by carrier traps in the nearby substrate.
Tunability with the Fermi level, close to and far from the charge neutrality point, and temperature dependence 
of the noise amplitude are clear fingerprints of the underlying material-inherent processes. 
Our results suggest a roadmap for the analysis of decoherence sources in the
implementation of coherent  devices by hybrid nanostructures.
\end{abstract}

Graphene Josephson junctions (GJJ) in the regime of ballistic transport emerged in the last few years
as unique hybrid systems allowing investigation of fundamental quantum phenomena related to 
proximitized superconductivity in a two-dimensional (2D) material. 
High-quality graphene-superconductors heterostructures with clean interfaces, realized by encapsulating graphene 
in hexagonal boron nitride (hBN) with one dimensional edge contacts to superconducting leads, allowed the observation 
of ballistic transport of Cooper pairs over micron scale lengths, of gate-tunable supercurrents that persist at large  
parallel magnetic fields~\cite{calado_natnanotech_2015,benshalom_natphys_2016,borzenets_prl_2016}
and of different features of Andreev physics in two dimensions
\cite{allen_natphys_2016,amet_science_2016,bretheau_natphys_2017}.
Short ballistic-GJJ, with junction channel length much shorter than the superconducting coherence length, 
are characterized by a strongly non-sinusoidal current-phase relation (CPR) whose skewness and critical
current depend on gate voltage and temperature~\cite{titov_prb_2006,blackschaffer_doniach_2008,blackschaffer_linder_2010,
Hagymasi_PRB_2010,takane_jpsj_2011,takane_jpsj_2012}. 
Experimental evidences of strong Josephson coupling in planar ballistic GJJ have been recently reported  
 \cite{english_prb_2016,nanda_nanol_2017,park_prl_2018}.
Envisioned potential applications of GJJ range from ultrasensitive magnetometers and voltmeters to digital logic circuits. 
Very recent experimental studies integrated graphene-based van der Waals heterostructures into circuit quantum electrodynamics systems~\cite{kroll_natcomm_2018,steele_natcomm_2018,wang_natnanotech_2018}.
Spectroscopy and coherent quantum control in a graphene-based "gatemon" \cite{wang_natnanotech_2018}, 
together with microwave performances \cite{kroll_natcomm_2018} and resilience to strong magnetic fields \cite{steele_natcomm_2018}
make short ballistic-GJJs a promising platform for the implementation of coherent quantum circuits in hybrid architectures,
opening up a promising tool for topological quantum computing. 
Understanding material-inherent microscopic noise sources possibly limiting the phase-coherent behavior of GJJ-based quantum circuits represents an essential, still unexplored, prerequisite. 
Recently, indications of the possible presence of spurious two-level systems embedded in the heterostructure
emerged as unexpected frequency components in Fourier transform of Ramsey fringes \cite{wang_natnanotech_2018} and
out-of-gap energy features in tunneling spectroscopy measurements \cite{wang_prb_2018}.

A specially relevant issue is understanding the impact on ballistic GJJs of fluctuations responsible for current 
noise with $1/f$ power spectrum, which is observed in a variety of graphene devices~\cite{balandin_natnano_2013}. 
Discovered about a century ago, low frequency noise with $1/f$ power spectrum is
still an intriguing phenomenon occurring in a variety of materials and over different scales. 
Investigation of decoherence due to $1/f$ noise in superconducting quantum devices based on 
conventional Josephson junctions provides relevant insights into microscopic noise  sources \cite{paladino_rmp_2014}.
This has allowed developing quantum control strategies to reduce its effects towards the realization of
efficient basic elements for quantum information purposes.

Although detrimental in many of its manifestations, $1/f$ noise offers also opportunities for materials characterization.
Graphene, with its inherent bi-dimensional nature, linear energy dispersion for electrons and holes and zero-energy band
gap, is a unique material in the context of $1/f$ noise which has been observed even in clean samples of suspended 
graphene and in graphene on hBN substrates \cite{kumar_nanolett_2016,kumar_apl_2015,stolyarov_apl_2015,kayyalha_apl_2015}. 
$1/f$ noise is in fact a versatile probe to study fluctuations affecting charge transport properties, as
density fluctuations, charge dynamics, dielectric screening, which cannot be directly accessed by resistivity measurements. 
Remarkably, because of their strongly non-sinusoidal CPR with gate voltage-tunable skeweness and critical current, 
ballistic GJJ are potentially flexible quantum probes of microscopic fluctuations underlying  transport 
2D materials.

In this work we show that  fluctuations with $1/f$ power spectrum of the critical current, $I_{\rm c}$, of a short ballistic 
GJJ directly probe carrier density fluctuations of the graphene channel due to charge traps in the nearby substrate. 
Tunability with the Fermi level,
close to and far from the charge neutrality point, and temperature dependence of the noise amplitude are clear
fingerprints of the underlying material-inherent processes. 
Noise results from proximitized superconductivity 
of a normal metal, a mechanism peculiar of GJJs. Instead, in conventional Josephson junctions 
switching charge traps in the insulating barrier randomly block tunneling channels thus
modulating the junction area and inducing $1/f$ critical current noise
\cite{paladino_rmp_2014}.Our results provide also relevant figures of merit in view  of the  implementation of coherent quantum circuits in hybrid architectures.

In a short ballistic GJJ a dissipationless supercurrent flows in equilibrium through the proximitized normal
metal  region.  
The coherent flow of Cooper pairs through graphene is due to successive Andreev reflections
at the normal metal-superconducting  interfaces. 
In the ballistic limit, where the junction channel length $L$ is much shorter than the mean free path
$l_{\rm mfp}$, well-defined Andreev bound states are formed inside
the superconducting gap, $\Delta \equiv \Delta(T)$. The corresponding energies depend on the phase difference $\phi$ of the superconducting
order parameters on the two sides of the junction. Each Andreev level with energy $\varepsilon(\phi)$, carries 
a supercurrent  $(1/\Phi_0) \partial \varepsilon(\phi)/\partial \phi$, where $\Phi_0=\hbar/2e$ is the flux quantum. 
In the short junction limit ($L \ll \xi, W$, where $\xi= \hbar v_{\rm D}/\Delta$ is the superconducting coherence length, 
$W$ is the channel width, and $v_{\rm D}$ is the graphene monolayer Fermi velocity $v_{\rm D}\approx 10^6~{\rm m/s}$)  the supercurrent is mediated by  a single bound state, $\varepsilon(q_n,\phi)$, per 
transversal mode $q_n=(n+1/2)\pi/W$. This mechanism results in a non-sinusoidal CPR \cite{titov_prb_2006,blackschaffer_doniach_2008,blackschaffer_linder_2010,Hagymasi_PRB_2010}.
 Recent measurements of the CPR in the ballistic regime revealed a gate-tunable skewness, sensitive to the 
junction length and to the nature of the superconducting-graphene interface~\cite{english_prb_2016,nanda_nanol_2017}. 
The maximal supercurrent, $I_{\rm c}$, depends on the doping and it is nonvanishing even at the Dirac point, despite of 
the zero carrier concentration resulting from the linear dispersion of graphene. 
Gate voltage tunability of the critical current is a relevant feature for GJJ applications. Ultimately, it originates 
from the dependence of the Andreev bound states on the Fermi level, 
related in turn to the actual carrier density. As a consequence, fluctuations of carrier density in the 
graphene insert are responsible for fluctuations of Andreev levels manifesting themselves as noise in the critical
current of the ballistic GJJ. 

A number of investigations on $1/f$ current (or equivalently resistance) noise in mono-layer graphene~\cite{balandin_natnano_2013}, and recently in  graphene tunnel junctions\cite{puczk_acsnano_2018}, pointed out 
the relevant  role of carrier density fluctuations due to  charge trapping and release processes between graphene 
and carrier traps in the underlying substrate. This noise mechanism, typical of conventional semiconducting FET,
is commonly described by the McWorther  model~\cite{mcwhorter_1957}. 
Each trap can be empty  or occupied by an electron, and it randomly switches between these two states. 
Typical switching times between the two states are much longer that the relaxation time of the crystal,
thus trapping-recombination traces are modeled as  Markovian random telegraph processes.
A  spatially uniform distribution of independent generation-recombination centers determines a 
logarithmic distribution of the switching rates, $1/\tau$, of the noise sources in the interval 
$[1/\tau_{max},1/\tau_{min}]$. This yields $1/f$ noise spectrum in the same frequency range~\cite{mcwhorter_1957,dutta_rmp_1981,weissman_rmp_1988,kogan_book,balandin_natnano_2013,paladino_rmp_2014},
the actual low-frequency cut-off $1/\tau_{max}$ being in practice hardly detectable.  
This is also the basis of our description of critical current noise in short ballistic GJJs.  

\begin{figure}[ht]
\centering
\begin{overpic}[width=0.59\columnwidth]{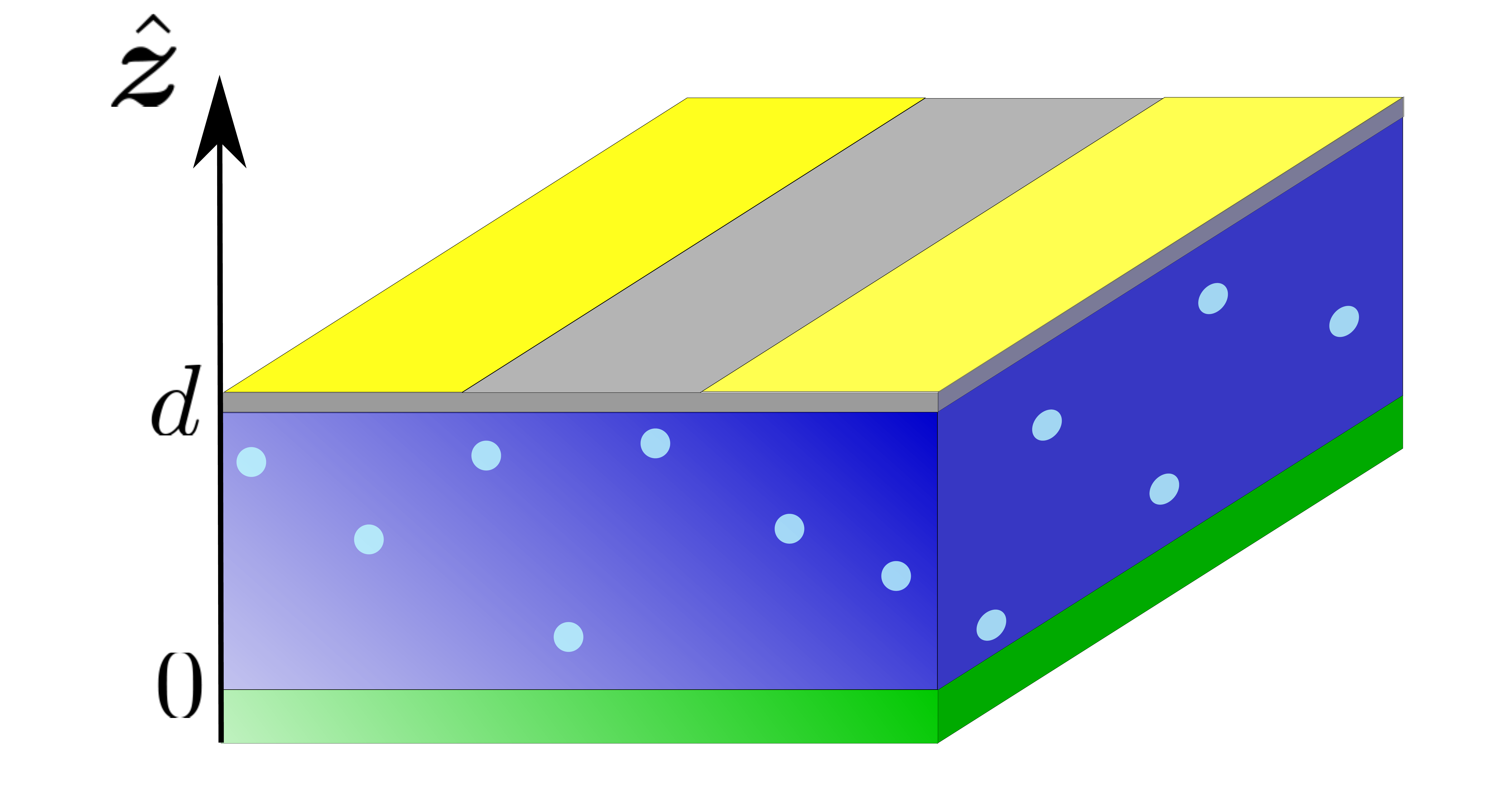}\put(2,51){\normalsize (a)}\end{overpic}
\begin{overpic}[width=0.39\columnwidth]{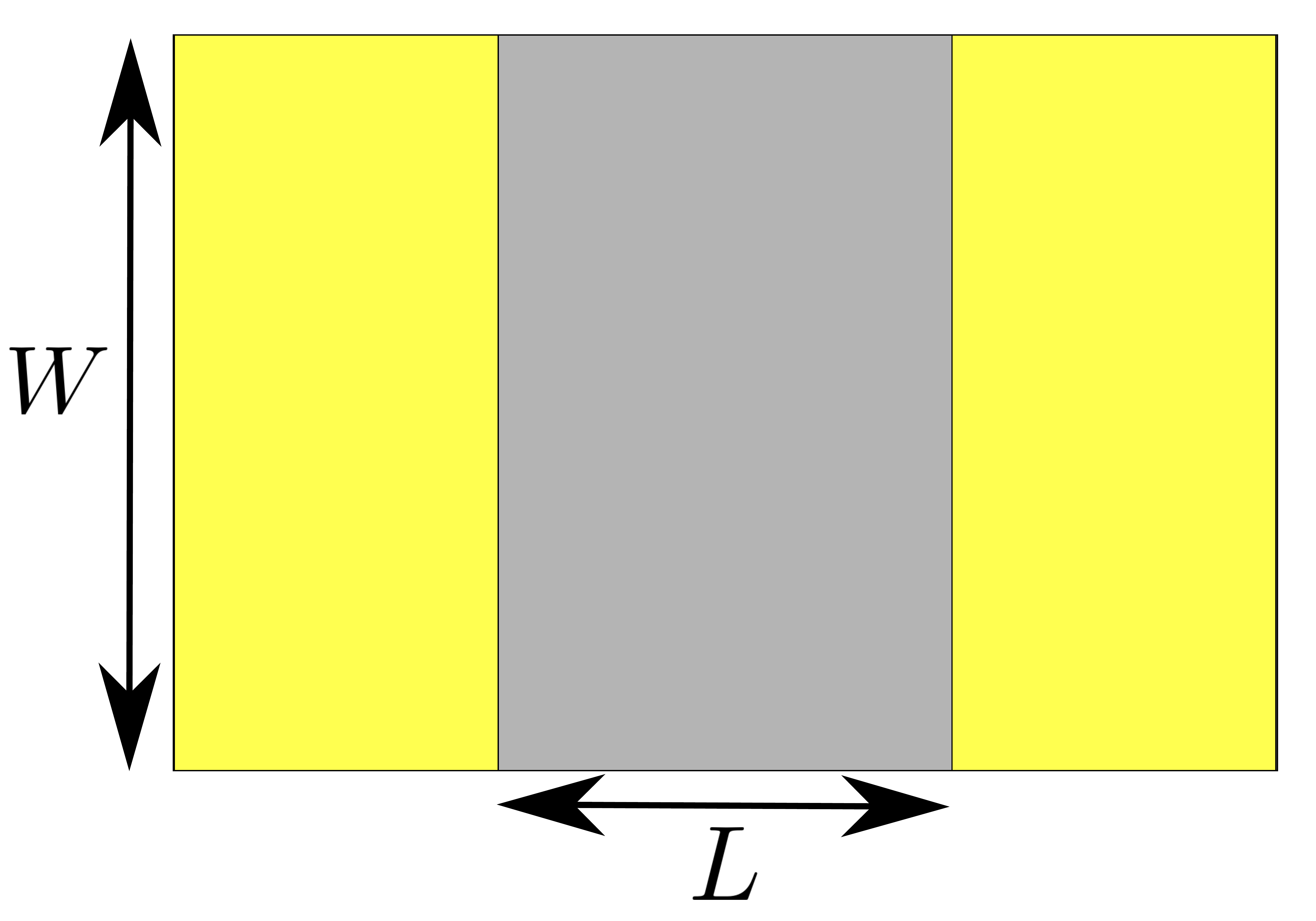}\put(2,78){\normalsize (b)}\end{overpic}\\
\caption{Schematic of the device. Panel (a) displays the side view, from bottom to top there are a metal gate
(green), a substrate (blue), a monolayer graphene  (gray) and two superconducting electrodes (yellow). 
Electron traps are represented with cyan circles randomly distributed inside the oxide substrate.
Panel (b) displays the top view, gray region represents the stripe in normal phase and yellow sides are the regions covered by superconductors.
Here, $L$ represents the junction channel and $W$ is the length of the device along the invariant direction.
}
\label{fig:scheme}
\end{figure}

{\bf Results}\\
{\bf Model.} The system considered in this work is schematically shown in  Fig.~\ref{fig:scheme}. A graphene layer (gray), partially covered by two superconducting electrodes (yellow),
is deposited on a substrate (blue) under which a metal gate (green) allows electrical tuning of the doping level in graphene. Carrier traps, randomly distributed in the oxide, are represented by cyan circles.

We model the ballistic GJJ within the  Dirac-Bogoliubov-de Gennes approach
where  superconducting metal stripes induce on the underlying graphene layer very large doping
and superconductivity by proximity effect \cite{titov_prb_2006,takane_jpsj_2011,takane_jpsj_2012}.
In the short junction limit the supercurrent is expressed as
\begin{equation}\label{eq:superI}
 I(\phi)\equiv-\frac{4 e}{\hbar} \sum^\infty_{n=0} \tanh\left[ \frac{\varepsilon(q_n,\phi)}{2 k_{\rm B} T} \right] \frac{\partial \varepsilon(q_n,\phi)}{\partial \phi}~,
\end{equation}
where the Andreev eigenenergies depend on the phase difference $\phi$ and on the normal-state transmission amplitude $\tau(q_n)$ as 
\begin{eqnarray}
\label{eq:epsn}
\varepsilon(q_n,\phi) &=& \Delta \sqrt{ 1-\tau(q_n) \sin^2(\phi/2)}~, \\
\tau(q_n) &=& \frac{k_{\rm F}^2-q_n^2}{k_{\rm F}^2-q_n^2 \cos^2(\sqrt{k_{\rm F}^2-q_n^2}L)}~.
\end{eqnarray}
Here $k_{\rm F}=\mu_0/(\hbar v_{\rm D})$ is the Fermi wavenumber expressed  in terms of the  Fermi level $\mu_0$ 
and the graphene monolayer Fermi velocity $v_{\rm D}$.
For a wide and short normal region (see  Fig.~\ref{fig:scheme}~(b)), $L \ll W, \xi$, the summation in Eq.~(\ref{eq:superI}) can be replaced by an integral. 
Recently,  ballistic devices in this limit have been experimentally realized by using graphene encapsulated in hBN~\cite{borzenets_prl_2016,nanda_nanol_2017,park_prl_2018}. 
Maximization of Eq.~(\ref{eq:superI}) with respect to $\phi$ gives the junction's critical current, $I_{\rm c}(\mu_0,T)$. 
Because of the dependence of the Fermi level on the carrier density, both the CPR and the critical current 
 are tunable with the gate voltage.  
At zero temperature $I_{\rm c}$ at the charge neutrality point (CNP), $\mu_0=0$,
is approximately given by  $1.33 e \Delta_0 W/(\pi \hbar L)$, where $\Delta_0$ is the zero-temperature superconducting gap, $\Delta_0=\Delta(T=0)$~\cite{titov_prb_2006}.
The finite supercurrent, in the absence of free carriers in the graphene channel, is due to evanescent modes. With increasing values of doping level $|\mu_0|$ the critical current increases due to the contribution of propagating modes,  
independently of the sign of carriers because of electron-hole symmetry. Its dependence on the Fermi level 
changes from parabolic close to the CNP to the linear asymptote, 
$1.22 e \Delta_0 W |\mu_0|/(\pi \hbar^2 v_{\rm D})$, for large doping  $|\mu_0|\gg \hbar v_{\rm D}/L$.
Small amplitude Fabry-Perot oscillations appear for finite doping due to the interference of reflected carriers 
at the graphene-superconductor interfaces, characteristic of the ballistic regime~\cite{titov_prb_2006}.

{\bf Critical current noise.}
Whenever the Fermi level deviates from the equilibrium value $\mu_0$, the critical current
 fluctuations  can be approximated as
\begin{equation}
 \delta I_{\rm c}(t) = I_{\rm c}(t) -\langle I_{\rm c}(t) \rangle~,
\end{equation}
where
\begin{equation}\label{eq:expgen}
I_{\rm c}(t) \approx  I_{\rm c}(\mu_0) + \frac{d I_{\rm c}}{d \mu_0} \delta \mu(t)+  
 \frac{1}{2}  \frac{d^2 I_{\rm c}}{d \mu_0^2}[\delta \mu(t)]^2~.
\end{equation}
Close to the charge neutrality point, where the critical current first derivative vanishes, the dominant contribution to  current fluctuations  is
quadratic in the fluctuations of the Fermi level, whereas for large dopings the leading contribution is linear in  $\delta \mu(t)=\mu(t)-\langle \mu(t)\rangle=\mu(t)-\mu_0$.
In the following, we relate fluctuations of the Fermi level to carrier density fluctuations due to trapping/recombination processes within the McWorther model and evaluate the critical current power spectrum
\begin{equation}\label{eq:SIc_def}
{\cal S}_{ I_{\rm c}} (\omega)\equiv \int_0^\infty \frac{d t}{\pi}\cos(\omega t) \langle \delta  I_{\rm c}(t) \delta  I_{\rm c}(0) \rangle~,
\end{equation}
where the current-current correlation function is written in terms of second and  higher order correlators of $\delta \mu(t)$
\begin{eqnarray}
\langle \delta  I_{\rm c}(t) \delta  I_{\rm c}(0) \rangle &=& 
 \left[ \frac{d I_{\rm c}(\mu_0)}{d \mu_0} \right]^2  \langle \delta \mu(t) \delta \mu(0)\rangle \nonumber \\
&+& \frac{1}{2}\frac{d I_{\rm c}(\mu_0)}{d \mu_0}\frac{d^2 I_{\rm c}(\mu_0)}{d \mu_0^2}
 \langle [\delta \mu(t)]^2 \delta \mu(0)+\delta \mu(t) [\delta \mu(0)]^2\rangle \nonumber \\
&+& \frac{1}{4}\left[\frac{d^2 I_{\rm c}(\mu_0)}{d \mu_0^2} \right]^2 \left\{ \langle[ \delta \mu(t) ]^2[ \delta \mu(0)]^2\rangle -  \langle[ \delta \mu(t) ]^2\rangle 
\langle[ \delta \mu(0)]^2\rangle
\right\}~.
\label{Ic-mu}
\end{eqnarray}
In our model, fluctuations of the Fermi level stem from carriers trapped in the substrate.
Charge traps are randomly distributed in the substrate beneath the graphene 
layer~\cite{hooge_rps_1981,fernandez_prb_2007}, as sketched in  Fig.~\ref{fig:scheme}.
Charge carrier tunneling between the graphene electron channel and the substrate traps
induces a fluctuating voltage~\cite{fusah_ieee_1972},  $V_{\rm T} (t)$, 
which contributes to the (fixed) voltage drop between the metal gate and the graphene layer,  $V_{\rm G}$, 
\begin{equation}\label{eq:Vg}
 V_{\rm G} = \frac{W_{\rm f}}{e}+ \frac{4 \pi e d n(t)}{\epsilon_{\rm r}} + \frac{\mu(t)}{e}  + V_{\rm T} (t)~,
 \end{equation}
where $W_{\rm f}$ is the work function difference between the gate and graphene. 
The other two terms are the geometric and quantum capacitance contributions due to  charge carriers in the
graphene layer,
$d$ and $\epsilon_{\rm r}$ being respectively the width and the dielectric constant of the substrate, and $n(t)$ the 
instantaneous carrier density in graphene. 
The equilibrium carrier density $n_0$ is related to the Fermi level $\mu_0$, in particular
at zero temperature $\mu_0= \hbar v_{\rm D} \, \sqrt{\pi |n_0|}$~\cite{giulianivignale_book}. 
Being a disordered system, charge traps are spatially randomly distributed in the substrate layer 
and have an unknown distribution in energies $\epsilon$ (with respect to the charge neutrality point, $\mu_0=0$).  
If we assume that the spatial distribution of carrier traps is quasiuniform along the $\hat{\bm x}$ and $\hat{\bm y}$ directions~\cite{brews_jap_1974}, 
the voltage drop $V_{\rm T} (t)$ can be written as
\begin{equation}\label{eq:Vt}
V_{\rm T} (t) = \frac{4 \pi e}{\epsilon_{\rm r}} \int \frac{d {\bm r} }{L W}\int_0^d dz z
\int_{-\infty}^\infty d \epsilon  \, {\cal N}_{\rm T} (\epsilon,\bm R,t)~,
\end{equation}
where ${\bm R}=({\bm r},z)$ and ${\cal N}_{\rm T}(\epsilon,\bm R,t)$ denotes the density of populated traps  
per unit volume and energy.
In equilibrium, it reads
${\cal N}_{\rm T0} (\epsilon,\bm R)= f_{\rm D}(\epsilon-\mu_0)   {\cal D}(\epsilon,\bm R)$,
where ${\cal D}(\epsilon,\bm R)$ is the number of trap states per unit of energy and volume whose
occupation probability is given by  the Fermi distribution $f_{\rm D}(x)=1/[e^{x/(k_{\rm B} T)}+1]$.
Since the time scale of fluctuations of carriers in graphene is much shorter than the time scale of the charge fluctuations in the traps~\cite{kogan_book}, we assume that 
charge carriers (as well as Fermi level) in graphene adjust instantaneously to  fluctuations of the trapped carriers
entering  $\delta V_{\rm T}(t)$. 
Under these conditions, expansion of Eq.~(\ref{eq:Vg}) around the equilibrium values gives
\begin{equation}\label{eq:mut_ast}
\delta \mu(t)=-e \frac{C_{\rm g}}{C_\parallel} \delta V_{\rm T}(t)~,
\end{equation}
where $C_\parallel=C_{\rm g}+C_{\rm Q}$, $C_{\rm g}\equiv\epsilon_{\rm r}/(4 \pi d)$ is the geometric capacitance, $C_{\rm Q}$ is the quantum capacitance
\begin{equation}
 C_{\rm Q} \equiv e^2 \frac{d n_0}{d \mu_0}=\frac{2 e^2  }{\pi \hbar^2 v_{\rm D}^2}   k_{\rm B} T
 \ln\left[2+2\cosh\left( \frac{\mu_0}{k_{\rm B} T} \right) \right]~,
\end{equation}
and  $ \delta V_{\rm T}(t)$ represents the  deviations of the trap voltage drop from the equilibrium value due to population fluctuations of
the trap density with respect to ${\cal N}_{{\rm T}0}(\epsilon,\bm R)$
\begin{equation}
\label{eq:deltaVT}
\delta V_{{\rm T}}(t) = \frac{4 \pi e}{\epsilon_{\rm r}} \int \frac{d {\bm r} }{L W}\int_0^d dz z   \int_{-\infty}^\infty d \epsilon  
\delta {\cal N}_{\rm T}(\epsilon,{\bm R},t )~,
\end{equation}
where $\delta {\cal N}_{\rm T}(\epsilon,{\bm R},t )=   {\cal N}_{\rm T} (\epsilon,\bm R,t) -  {\cal N}_{\rm T0} (\epsilon,\bm R)$
can be expressed as
\begin{equation}\label{eq:nt_random}
\delta {\cal N}_{\rm T}(\epsilon,{\bm R},t )\equiv \sum_{i}\delta(\epsilon-\epsilon_i) \delta({\bm R}-{\bm R}_i) [X(i,t)- f_{\rm D}(\epsilon-\mu_0)]~,
\end{equation}
and $X(i,t)$ is a random telegraph process,
being one (zero) when the trap $i$ is filled (empty) \cite{kogan_book}.
Switching between the occupied/empty state of trap $i$ occurs with a rate depending on the trap position along the 
direction perpendicular to the graphene layer~\cite{mcwhorter_1957,balandin_natnano_2013}
\begin{equation}\label{eq:tunnel}
 \gamma(z)=\gamma_0 \exp(-|z-d|/\ell)+\gamma_0^\prime \exp(-|z|/\ell^\prime)~,
\end{equation}
where we distinguish tunneling processes related to the graphene channel, characterized by $\gamma_0$ and $\ell$, and tunneling process related to the gate channel, characterized by $\gamma_0^\prime$ and $\ell^\prime$. 
Typical orders of magnitude of the tunneling parameters are $\gamma_0, \gamma_0^\prime \sim 10^{10}~{\rm s}^{-1}$ and $\ell, \ell^\prime \sim 1~$\AA~\cite{balandin_natnano_2013}. 
The Fermi level correlators entering $I_{\rm c}$'s fluctuations Eq.(\ref{Ic-mu}) are therefore related to
correlators of various orders of the population of traps.
Exploiting  Markovianity, assuming that traps are uncorrelated  and $\langle X(i,t) \rangle=f_{\rm D}(\epsilon_i- \mu_0)$, 
the correlators up to the $4$-th order in the population fluctuations of trapped electron density are written as
\begin{subequations}\label{eq:correlators}
\begin{equation}
 \begin{aligned}
\langle \delta {\cal N}_{\rm T}(\epsilon_1,{\bm R}_1,t_1 ) \delta {\cal N}_{\rm T}(\epsilon_0,{\bm R}_0,t_0 ) \rangle &=
\delta({\bm R}_1 -{\bm R}_0) \delta(\epsilon_1-\epsilon_0)  {\cal D}(\epsilon_0,{\bm R}_0) 
f_{\rm D}(\epsilon_0- \mu_0)\\
&\times[1-f_{\rm D}(\epsilon_0- \mu_0)] \exp[-\gamma(z_0)(t_1-t_0)]~,
\end{aligned}
\end{equation} 
\begin{equation}
 \begin{aligned}
\langle  \prod^2_{k=0}  \delta {\cal N}_{\rm T}(\epsilon_k,{\bm R}_k,t_k )  \rangle   &=
\prod^{1}_{k=0} \delta({\bm R}_{k+1} -{\bm R}_k) \delta(\epsilon_{k+1}-\epsilon_k)
  {\cal D}(\epsilon_0,{\bm R}_0)  f_{\rm D}(\epsilon_0- \mu_0)\\
 &\times  [1-f_{\rm D}(\epsilon_0- \mu_0)]  [1-2 f_{\rm D}(\epsilon_0- \mu_0)]  \exp[-\gamma(z_0)(t_2-t_0)]~, 
\end{aligned}
\end{equation} 
\begin{equation}
 \begin{aligned}
\langle  \prod^3_{k=0}  \delta {\cal N}_{\rm T}(\epsilon_k,{\bm R}_k,t_k )  \rangle   &=
\langle \delta {\cal N}_{\rm T}(\epsilon_3,{\bm R}_3,t_3 ) \delta {\cal N}_{\rm T}(\epsilon_2,{\bm R}_2,t_2 ) \rangle
\langle \delta {\cal N}_{\rm T}(\epsilon_1,{\bm R}_1,t_1 ) \delta {\cal N}_{\rm T}(\epsilon_0,{\bm R}_0,t_0 ) \rangle\\
&+\prod^{2}_{k=0} \delta({\bm R}_{k+1} -{\bm R}_k) \delta(\epsilon_{k+1}-\epsilon_k)
  {\cal D}(\epsilon_0, {\bm R}_0)  f_{\rm D}(\epsilon_0- \mu_0)\\
 &\times  [1-f_{\rm D}(\epsilon_0- \mu_0)]  [1-2 f_{\rm D}(\epsilon_0- \mu_0)]^2  \exp[-\gamma(z_0)(t_3-t_0)]~,
\end{aligned}
\end{equation}
\end{subequations} 
and $\langle  \delta {\cal N}_{\rm T}(\epsilon,{\bm R},t )  \rangle=0$ (see details in supplementary information).
By using Eq. (\ref{eq:mut_ast}) with (\ref{eq:deltaVT}) and the correlators  in Eqs.~(\ref{eq:correlators}),
considering that $d\ll \ell, \ell^\prime$ in the switching rates,  Eq.~(\ref{eq:tunnel}),
the critical current spectrum, Eq. (\ref{eq:SIc_def}), for frequencies $\omega \ll \gamma_0, \gamma_0^\prime$ takes  the  characteristic form 
$ {\cal S}_{ I_{\rm c}} (\omega) = {\cal A}_{ I_{\rm c}}/\omega$  with amplitude  
\begin{equation}\label{eq:SIc}
{\cal A}_{ I_{\rm c}} =
 \Bigg[
  \Bigg(  \frac{d I_{\rm c}}{d \mu_0}  \Bigg)^2 F_0 
 - \Bigg(  \frac{d I_{\rm c}}{d \mu_0}  \Bigg) \Bigg(  \frac{d^2 I_{\rm c}}{d \mu_0^2}  \Bigg) \varepsilon_{\rm Q} F_1  
 +\Bigg(  \frac{d^2 I_{\rm c}}{d \mu_0^2}  \Bigg)^2   \frac{\varepsilon_{\rm Q}^2}{4}  F_2
\Bigg]
\varepsilon_{\rm Q}^2 \frac{L W \ell }{2}~,
\end{equation}
where  $\varepsilon_{\rm Q}= e^2/(C_\parallel L W)$ and
\begin{equation}\label{eq:Fj}
 F_j \equiv  \int_{-\infty}^\infty d \epsilon  {\cal D}(\epsilon)
 f_{\rm D}( \epsilon -\mu_0)[1-f_{\rm D}( \epsilon -\mu_0)][1-2f_{\rm D}( \epsilon -\mu_0)]^j~,
\end{equation}
having assumed that the density of trap states does not depend on ${\bm R}$ and indicated it as
 ${\cal D}(\epsilon)$.
The critical current power spectrum with amplitude given by Eq.~(\ref{eq:SIc}) is the main result of this work.
The three contributions entering the noise amplitude arise from correlators of the trapped electron density 
populations of different orders.
The term proportional to $F_0$ derives from second order correlator, while the terms in $F_1$ and $F_2$ derive from 
correlators of the third and fourth order (see supplementary information).
Their contribution to the noise
amplitude depends on the doping level, $\mu_0$, and on temperature. In the undoped case, being 
$d I_{\rm c}/ d \mu_0 |_{\mu_0=0}=0$, the spectrum reduces to
\begin{equation}\label{eq:SIc_undoped}
{\cal S}_{ I_{\rm c}} (\omega)\Big|_{\mu_0=0}=
\Bigg(  \frac{d^2 I_{\rm c}}{d \mu_0^2}  \Bigg)^2   
\frac{e^8 \ell  F_2}{8 C_{\parallel}^4(LW)^3} \frac{1}{\omega}~,
\end{equation}
for large doping instead 
\begin{equation}\label{eq:SIc_bdc}
 {\cal S}_{ I_{\rm c}} (\omega) \approx \Bigg(  \frac{d I_{\rm c}}{d \mu_0}  \Bigg)^2 \frac{e^4 \ell F_0}{2 C_{\rm Q}^2LW} \frac{1}{\omega}~.
\end{equation}
Thus by tuning the doping level, the GJJ's critical current spectrum probes either the power spectrum
(large doping) or higher order correlators  of the trapped electron density population.
At the CNP the  $I_{\rm c}$ spectrum is a measure of the fourth order correlator.
These correlators sensitively depend on the trap energy distribution ${ \cal D}_\Gamma(\epsilon)$, entering the functions $F_j$s, Eq. (\ref{eq:Fj}). 

In our phenomenological model we consider a Lorentzian distribution around a central energy $\epsilon_{\rm T}$ and with width $\Gamma$  
\begin{equation}\label{eq:Dlorentzian}
{ \cal D}_\Gamma(\epsilon) \equiv \frac{\rho_{\rm T}}{\pi} \frac{\Gamma}{(\epsilon-\epsilon_{\rm T})^2 + \Gamma^2}~.
\end{equation}
In the limit $\Gamma \to 0$ the distribution tends to a Dirac delta function $\rho_{\rm T} \delta(\epsilon-\epsilon_{\rm T})$, describing degenerate traps, whereas for large $\Gamma$ we model a uniform distribution,
 $\rho_{\rm T}/(\pi \Gamma)$. In these two limiting cases the power spectrum can be evaluated in analytic form (see supplementary information). 
From now on, in order to compare our results with realistic devices, we fix $d=0.1~\mu{\rm m}$,  $L=0.2~\mu{\rm m}$ and $W=3~\mu{\rm m}$. 
Moreover, we set the relative dielectric constant at $\epsilon_{\rm r}=4.4$ and  the gap energy 
at  $\Delta=0.1 \hbar v_{\rm D}/L$ which ensures the validity of the short junction limit, 
 $\xi \sim \hbar v_{\rm D}/\Delta \gg L$.

\begin{figure}[ht]
\centering
\begin{overpic}[width=0.49\columnwidth]{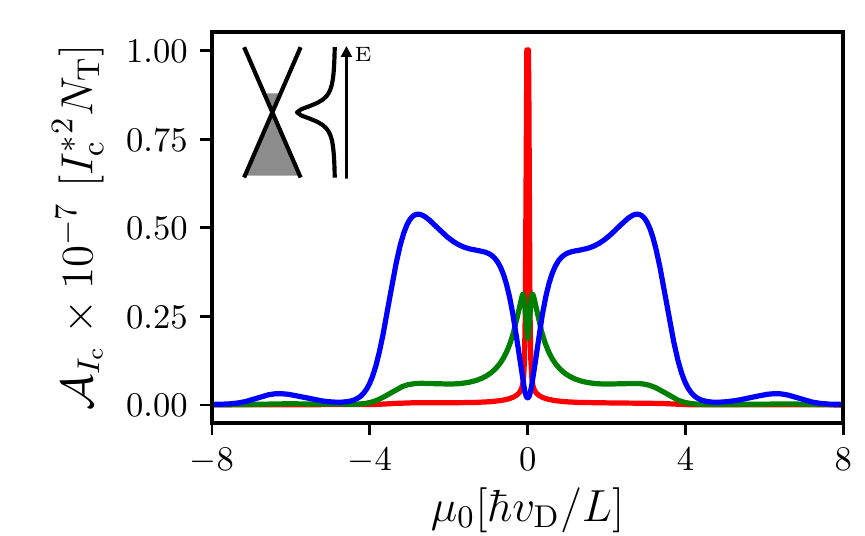}\put(-0.5,52){\normalsize (a)}\end{overpic}
\begin{overpic}[width=0.49\columnwidth]{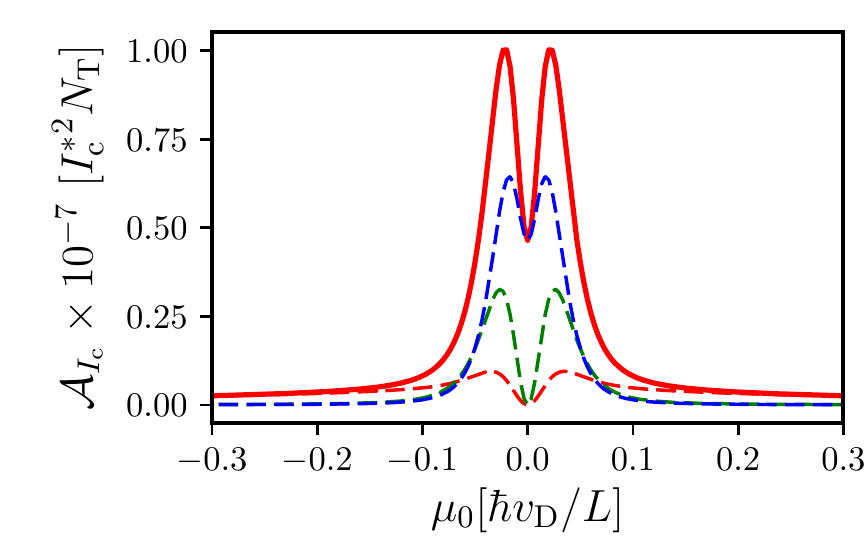}\put(-0.5,52){\normalsize (b)}\end{overpic}\\
\begin{overpic}[width=0.49\columnwidth]{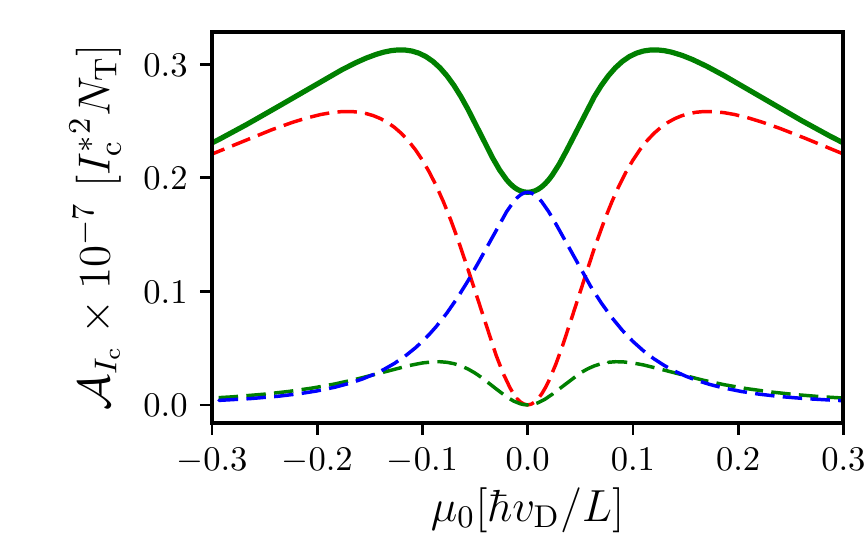}\put(-0.5,52){\normalsize (c)}\end{overpic}
\begin{overpic}[width=0.49\columnwidth]{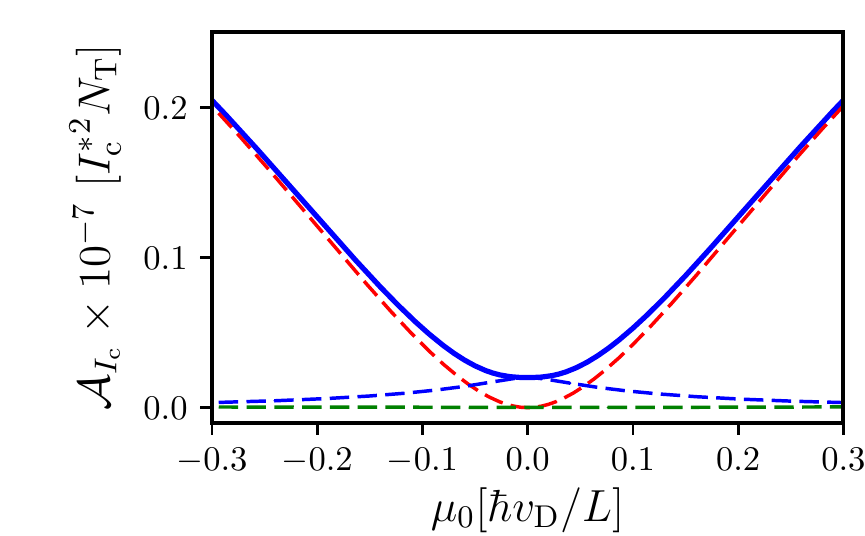}\put(-0.5,52){\normalsize (d)}\end{overpic}
\caption{Amplitude of the critical current spectrum, ${\cal A}_{I_{\rm c}}$, as a function of the doping level 
$\mu_0$ for trap energy distributions centered at the CNP, $\epsilon_{\rm T}=0$.
The amplitude is expressed in units of ${I_{\rm c}^\ast}^2 N_{\rm T}$ where $I_{\rm c}^\ast \equiv e \Delta W/(\hbar L)$  and $N_{\rm T}= \rho_{\rm T}  W L \ell$ is the number of traps in a slab of the substrate  of width $\ell$ under the graphene layer. Other parameters are $k_{\rm B} T=0.1 \Delta$ and $\Delta=0.1\hbar v_{\rm D}/L$.  
Colored lines correspond to different widths of the trap energy distribution $\Gamma$:
$\Gamma=0.01 \hbar v_{\rm D}/L$ (red solid line),  $\Gamma=0.1 \hbar v_{\rm D}/L$ (green solid line),  and $\Gamma= \hbar v_{\rm D}/L$ (blue solid line).
Panel (a) compares the amplitudes of the critical current power spectrum for the considered $\Gamma$. 
The inset shows a sketch of the graphene band with electronic states occupied up to a generic doping level (gray), 
and the Lorentzian trap energy distribution centered at the CNP. 
In panels (b), (c) and (d) the width is fixed to $\Gamma=0.01 \hbar v_{\rm D}/L$, $\Gamma= 0.1 \hbar v_{\rm D}/L$
and $\Gamma= \hbar v_{\rm D}/L$ respectively. Each panel compares the amplitude (solid line) with the corresponding correlators of  second order (red dashed line), third order (green dashed line) and fourth order  (blue dashed line) in proximity of $\mu_0=0$.}
\label{fig:SIc_mu_CNP}
\end{figure}

The dependence of the amplitude ${\cal A}_{I_{\rm c}}$ on the doping level 
is reported in  Fig.~\ref{fig:SIc_mu_CNP} for $T=0.1 \Delta/k_{\rm B}$ and trap energy distribution centered at 
the CNP, $\epsilon_{\rm T}=0$, for different widths $\Gamma$.
The noise amplitude is symmetric around the resonance condition, $\mu_0 =\epsilon_{\rm T} =0$.
For a narrow trap energy distribution, $\Gamma \ll \hbar v_{\rm D}/L $,  noise is non vanishing and takes large values
only for low doping.          
For a broader  trap energy distribution, the doping range where the amplitude is non vanishing increases 
and reflects $I_{\rm c}$'s Fabry-Perot oscillations, characteristic of the ballistic transport regime~\cite{titov_prb_2006}.
The behavior of ${\cal A}_{ I_{\rm c}}$ close to the CNP and the contributions from different correlators (dashed lines)
of the trapped electron density are reported in  Figs.~\ref{fig:SIc_mu_CNP}~(b), (c) and (d).
Correlators of orders larger than the second have a substantial impact on the critical current power spectrum in proximity of the CNP where it has an M-shaped trend independently of $\Gamma$.
For larger dopings the amplitude ${\cal A}_{I_{\rm c}}$ is dominated by the second order correlator, see Eq.~(\ref{eq:SIc_bdc}).

\begin{figure}[ht]
\centering
\begin{overpic}[width=0.49\columnwidth]{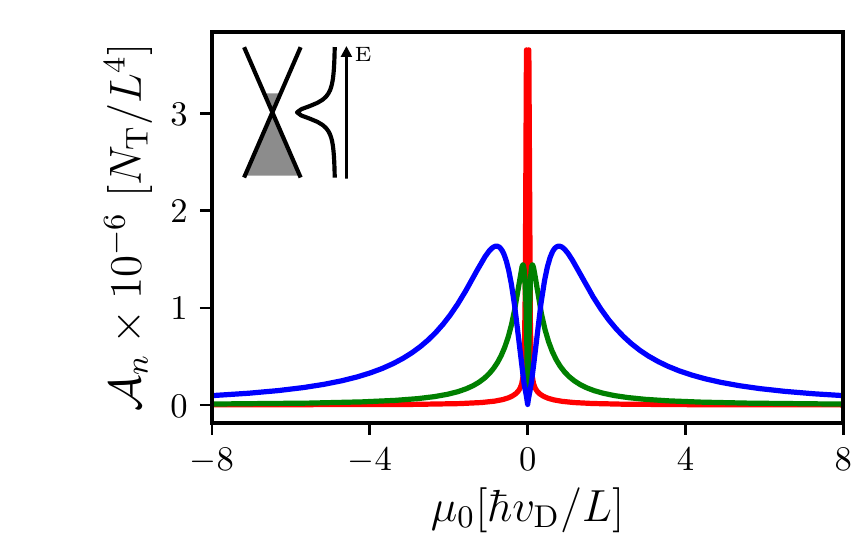}\put(-0.5,52){\normalsize (a)}\end{overpic}
\begin{overpic}[width=0.49\columnwidth]{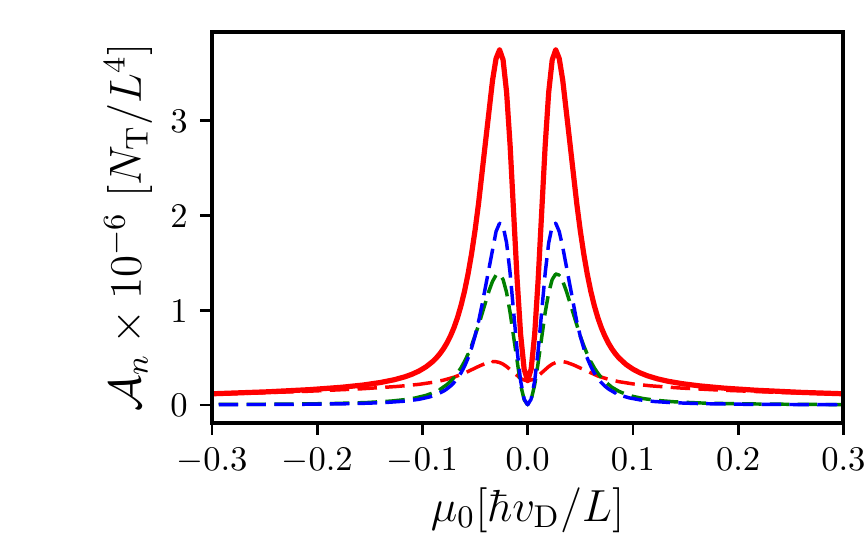}\put(-0.5,52){\normalsize (b)}\end{overpic}\\
\begin{overpic}[width=0.49\columnwidth]{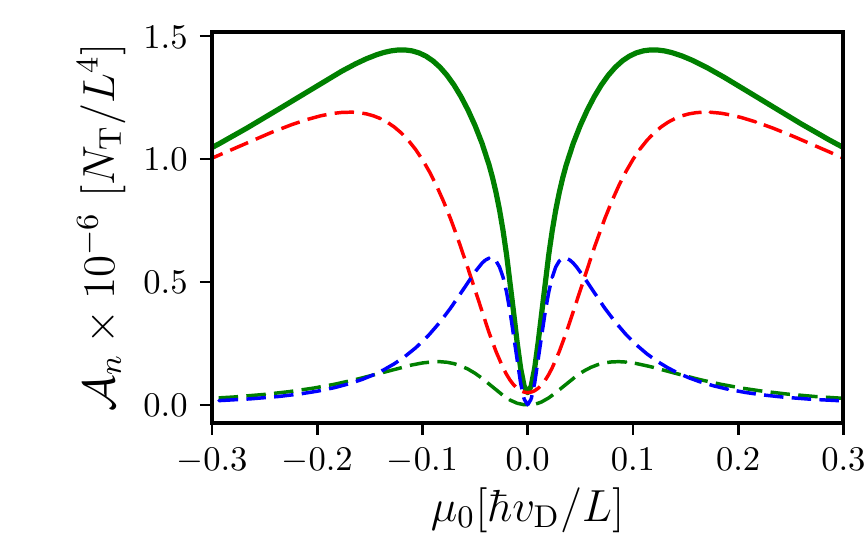}\put(-0.5,52){\normalsize (c)}\end{overpic}
\begin{overpic}[width=0.49\columnwidth]{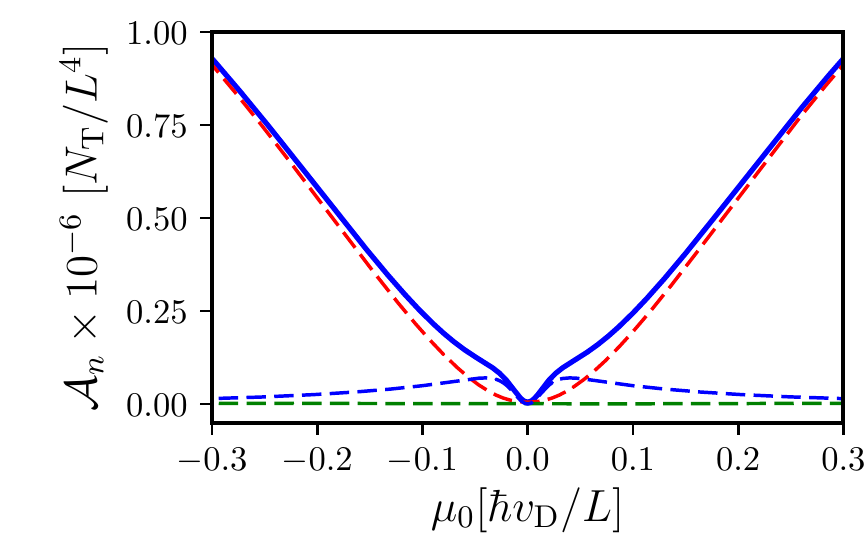}\put(-0.5,52){\normalsize (d)}\end{overpic}
\caption{Amplitude of the charge carrier density spectrum, ${\cal A}_{n}$, in units of $N_{\rm T}/L^4$, as a function of the doping level $\mu_0$, 
for $k_{\rm B} T=10^{-2} \hbar v_{\rm D}/L$ and trap energy distribution centered at the CNP, i.e. $\epsilon_{\rm T}=0$.
Colors correspond to different values of width $\Gamma$:
$\Gamma=0.01 \hbar v_{\rm D}/L$ (red solid line),  $\Gamma=0.1 \hbar v_{\rm D}/L$ (green solid line),  and $\Gamma= \hbar v_{\rm D}/L$ (blue solid line).
Panel (a): compares the amplitudes ${\cal A}_{n}$ for the considered $\Gamma$. 
The left-top inset shows a sketch of the electron structure with the shaded region below a generic doping level, 
and the Lorentzian trap energy distribution centered at the CNP. 
In panels (b), (c) and (c) the trap energy widhts are $\Gamma=0.01 \hbar v_{\rm D}/L$, $\Gamma= 0.1 \hbar v_{\rm D}/L$
and $\Gamma= \hbar v_{\rm D}/L$: each panel compares the amplitude (solid line) with the corresponding correlators of 
second (red dashed line), third (green dashed line) and fourth order  (blue dashed line) in proximity of $\mu_0=0$.
}
\label{fig:Sn_mu_CNP}
\end{figure}

\begin{figure}[ht]
\centering
\begin{overpic}[width=0.49\columnwidth]{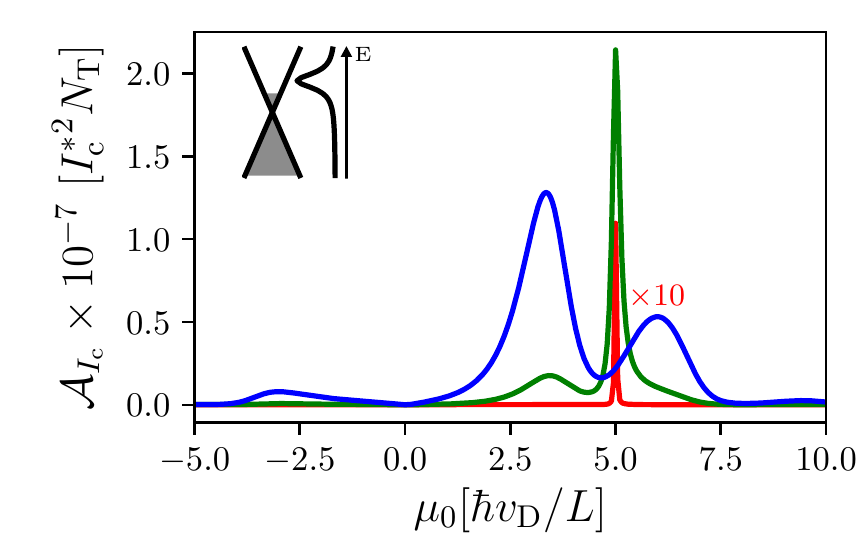}\put(2,52){\normalsize (a)}\end{overpic}
\begin{overpic}[width=0.49\columnwidth]{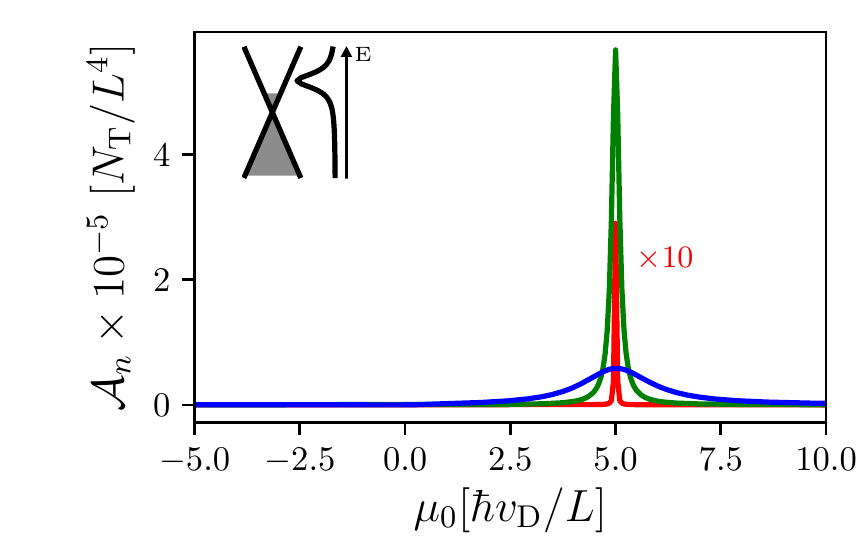}\put(2,52){\normalsize (b)}\end{overpic}
\caption{Critical current noise amplitude ${\cal A}_{I_{\rm c}}$,  in units of ${I_{\rm c}^\ast}^2 N_{\rm T}$ in panel (a),
and charge carrier noise amplitude ${\cal A}_n$, in units of  $ N_{\rm T}/L^4$ in panel (b), as a function of the doping 
level $\mu_0$. The trap energy distribution is centered at $\epsilon_{\rm T}=5 \hbar v_{\rm D}/L$ and widths
are $\Gamma=0.01 \hbar v_{\rm D}/L$ (red solid line, scaled to improve visibility),  $\Gamma=0.1 \hbar v_{\rm D}/L$ (green solid line),  and $\Gamma= \hbar v_{\rm D}/L$ (blue solid line). 
The left-top inset shows a sketch of the electron structure with the shaded region below a generic doping level, 
and the Lorentzian trap energy distribution centered in the conduction band. 
}
\label{fig:S_mu_bdc}
\end{figure}

{\bf Charge carrier density noise.}
Since both critical current and carrier density fluctuations are induced by  trapping-recombination processes, 
it is worth addressing also the carrier density spectrum. An independent detection of the two spectra could 
be used for a cross-check of the considered noise mechanism.  Carrier density fluctuations in GJJ could be
inferred from Hall voltage fluctuation measurements, similarly to the recent experiment on graphene~\cite{lu_prb_2014}.
Fluctuations of charge carrier density and of the doping level are related by 
\begin{equation}\label{eq:deltan}
 \delta n(t) = n(t) -\langle n(t) \rangle~,
\end{equation}
where
\begin{equation}
n(t) \approx  n_0 + \frac{C_{\rm Q}}{e^2 } \delta \mu (t)  + \frac{1}{2 e^2} \frac{d C_{\rm Q}}{d \mu_0}  \delta \mu (t)^2~,
\end{equation}
where $n_0$ represents the charge carrier density at equilibrium.
Using Eqs.~(\ref{eq:tunnel}) and (\ref{eq:correlators}), in the limit  $d\ll \ell, \ell^\prime$, 
the charge carrier density power spectrum for frequencies $\omega \ll \gamma_0, \gamma_0^\prime$, reads
\begin{equation}\label{eq:Sn_gen}
  {\cal S}_{ n}(\omega)=\Bigg[
C_{\rm Q}^2 F_0 
 - C_{\rm Q} \frac{d C_{\rm Q}}{ d \mu_0}  \varepsilon_{\rm Q} F_1  
 +\Bigg(  \frac{d C_{\rm Q}}{ d \mu_0}  \Bigg)^2   \frac{\varepsilon_{\rm Q}^2}{4}  F_2
\Bigg]
\varepsilon_{\rm Q}^2 \frac{L W \ell }{2 e^4} \frac{1}{\omega}  \equiv \frac{{\cal A}_n}{\omega}~.
\end{equation}
Remarkably, the two spectra have the same structure with $dI_{\rm c}/d \mu_0$ in ${\cal S}_{ I_{\rm c}} (\omega)$, Eq.~(\ref{eq:SIc}), replaced with the quantum capacitance, $C_{\rm Q}$, in ${\cal S}_n(\omega)$.
This quantity does not vanish at the CNP, where $C_{\rm Q}=C_{\rm Q} |_{\mu_0=0}= 4   \ln(2) e^2  k_{\rm B} T /(\pi \hbar^2 v_{\rm D}^2 )$, 
whereas $d C_{\rm Q}/ d \mu_0|_{\mu_0=0}=0$. Therefore, as a difference with 
$I_{\rm c}$'s spectrum, 
the charge carrier density spectrum at the CNP consists of the second order correlator in the 
trapped carriers density fluctuations,
\begin{equation}\label{eq:Sn_undoped}
{\cal S}_{n}(\omega)\Big|_{\mu_0=0}= \frac{{C_{\rm Q}}^2 \ell F_0 }{2 C_{\parallel}^2 L W } \frac{1}{\omega}~. 
\end{equation}
The dependence of the amplitude  ${\cal A}_{n}$ on the doping level is reported in  Fig.~\ref{fig:Sn_mu_CNP},
for the same temperature and trap energy distribution of  Fig.~\ref{fig:SIc_mu_CNP}. 
The amplitude ${\cal A}_{n}$ shows a M-shaped trend independently of $\Gamma$.
Whereas exactly at the CNP  ${\cal A}_n \propto F_0$, the impact of the correlators of orders larger than the second is substantial in proximity of the CNP where the size of the central dip at $\mu_0=0$ is sensitive to the trap energy distribution width $\Gamma$,
see  Figs.~\ref{fig:Sn_mu_CNP}~(b), (c) and (d). For larger doping the second order correlator dominates again
\begin{equation}\label{eq:Sn_bdc}
   {\cal S}_{ n}(\omega) \approx \frac{\ell F_0}{2 LW } \frac{1}{\omega}~.
\end{equation}
If the trap energy distribution instead of being centered at the CNP is centered in the conduction band 
both critical current and carrier density spectra are dominated by the second order correlators.
The amplitudes are asymmetric with respect to the resonance condition $\mu_0=\epsilon_{\rm T}$, due the 
electron-hole asymmetry, see   Fig.~\ref{fig:S_mu_bdc}.
Fabry-Perot oscillations in the amplitude of the current power spectrum appear clearly the larger is the width $\Gamma$ of the trap energy distribution,  Fig.~\ref{fig:S_mu_bdc}~(a).
The amplitude of the charge carrier density noise maintains instead a bell-shaped profile around $\epsilon_{\rm T}$, 
of larger width with broadening of the trap energy distribution, Fig.~\ref{fig:S_mu_bdc}~(b).

\begin{figure}[ht]
\centering
\begin{overpic}[width=0.49\columnwidth]{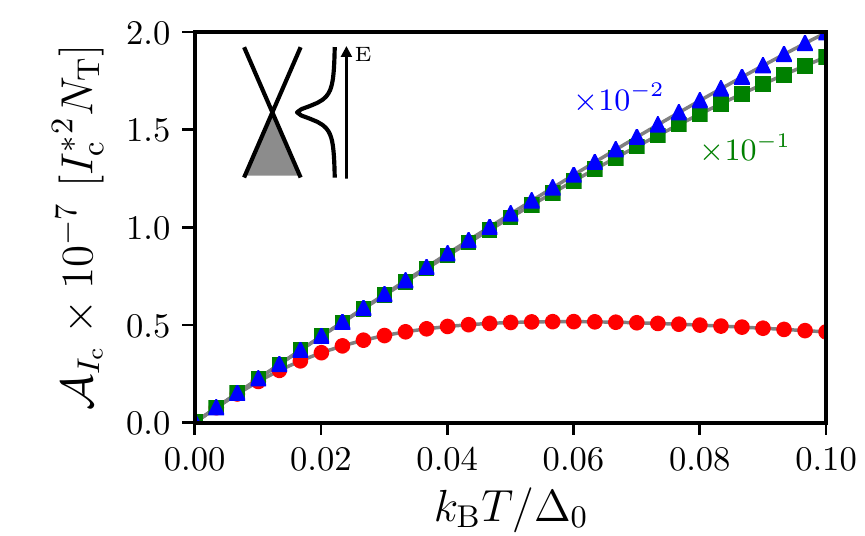}\put(-0.5,52){\normalsize (a)}\end{overpic}
\begin{overpic}[width=0.49\columnwidth]{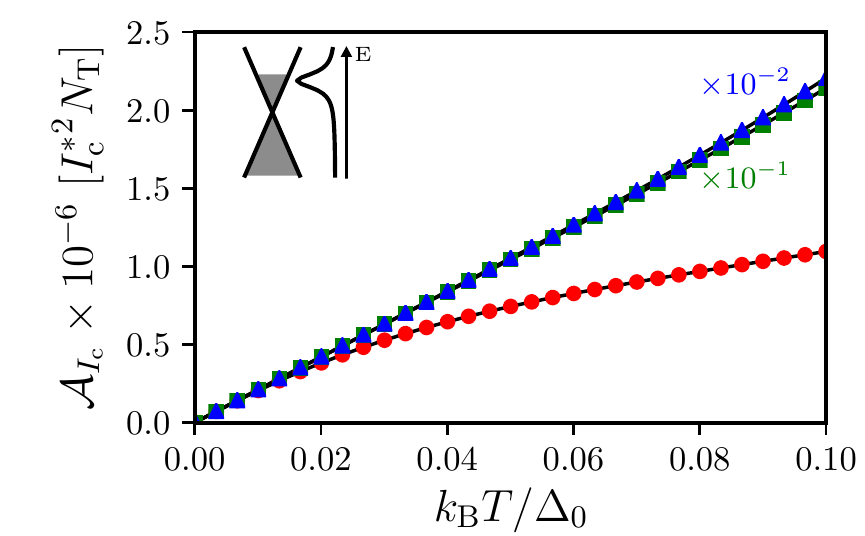}\put(-0.5,52){\normalsize (b)}\end{overpic}\\
\vspace{0.5em}
\begin{overpic}[width=0.49\columnwidth]{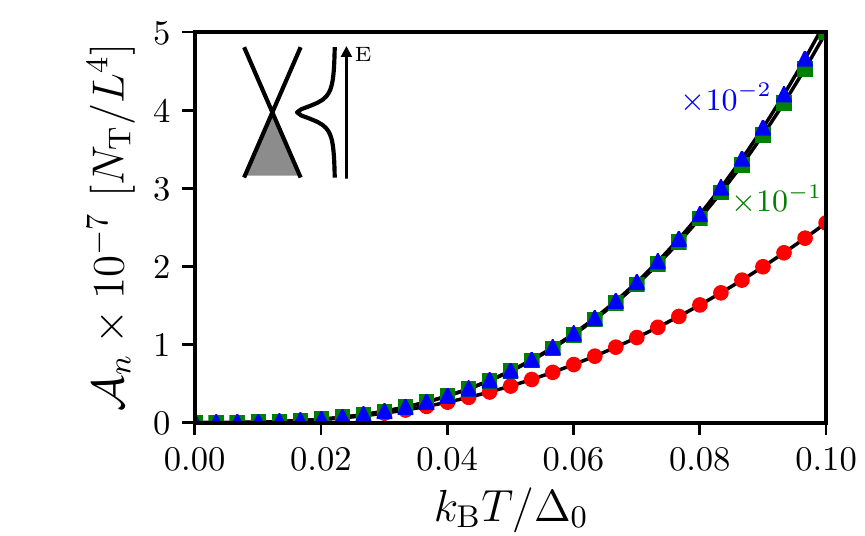}\put(-0.5,52){\normalsize (c)}\end{overpic}
\begin{overpic}[width=0.49\columnwidth]{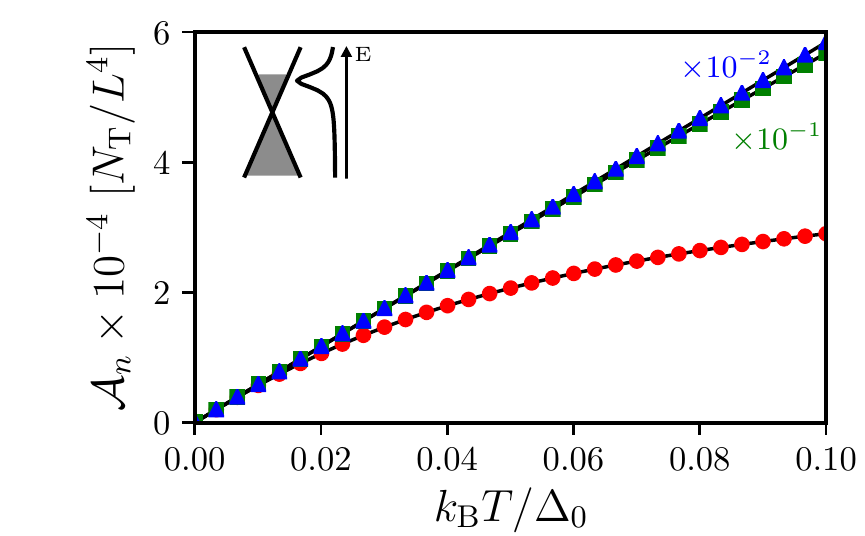}\put(-0.5,52){\normalsize (d)}\end{overpic}
\caption{Critical current noise amplitude ${\cal A}_{I_{\rm c}}$ (in units of ${I_{\rm c}^\ast}^2 N_{\rm T}$) 
(panels (a) and (c)) and carrier density noise amplitude ${\cal A}_n$ (in units of $ N_{\rm T}/L^4$) (panels (b) and (d))
 as a function of temperature for $\Delta_0=0.1 \hbar v_{\rm D}/L$ and fixed Fermi level $\mu_0=\epsilon_{\rm T}$. 
Top panels refer to a trap energy distribution at the CNP, $\epsilon_{\rm T}=0$,
in bottom panels the distribution is centered in the conduction band  $\epsilon_{\rm T}=5 \hbar v_{\rm D}/L$.
Different curves correspond to $\Gamma=0.01 \hbar v_{\rm D}/L$ (red circles), $\Gamma=0.1 \hbar v_{\rm D}/L$ (green squares),  
and $\Gamma= \hbar v_{\rm D}/L$ (blue triangles). Green and blue data have been scaled of  
 $0.01 \hbar v_{\rm D}/(L \Gamma)$ for the corresponding $\Gamma$ value.
In panel (a) the solid gray lines represent contributions from  correlators of the fourth order in the trapped carrier density fluctuations, see Eq.(\ref{eq:SIc_undoped}), while in 
in panels (b), (c) and (d) the solid black lines are contributions from correlators of the second order in the trapped carrier density fluctuations. In particular in panel (b) this corresponds to Eq. (\ref{eq:Sn_undoped}). 
Left-top insets show a sketch of the electron structure with a shaded region below the doping level placed at the center of trap energy distribution, and the Lorentzian trap energy distribution.}
\label{fig:S_T}
\end{figure}

{\bf Temperature dependencies.}
Charge trapping-release processes lead to peculiar temperature dependencies of both noise amplitudes. 
We consider low temperatures $k_{\rm B} T \ll \Delta_0$ and approximate  $\Delta (T) \approx \Delta_0$.
Figs.~\ref{fig:S_T}~(a) and (b) [(c) and (d)] display respectively the amplitudes  ${\cal A}_{I_{\rm c}}$ 
and ${\cal A}_n$  as a function of temperature, with the Fermi level and center of the trap energy distribution  
fixed at $\mu_0=\epsilon_{\rm T}=0$ [$\mu_0=\epsilon_{\rm T}=5 \hbar v_{\rm D}/L$].
At the CNP, the two amplitudes reflect the different temperature dependencies of the fourth and second order
correlator of the trapped carrier density fluctuations, as given by Eqs. (\ref{eq:SIc_undoped}) and (\ref{eq:Sn_undoped}).
The linear temperature behavior of  ${\cal A}_{ I_{\rm c}}$ for $T \to 0$ derives from the approximate form  
$F_2 \to \rho_{\rm T} k_{\rm B} T/(3 \pi \Gamma)$ approached in the limit  $\Gamma \to \infty$.
In Fig.~\ref{fig:S_T} amplitudes have been scaled of a factor $0.01 \hbar v_{\rm D}/(L \Gamma)$, 
so that in the linear temperature regime all curves superpose.
For larger temperatures $ T \lesssim \Gamma/ k_{\rm B}$, $F_2$ decreases monotonically. This regime is 
clearly visible in Fig.~\ref{fig:S_T}~(a) for the smallest $\Gamma$ value considered (red dots).
The carrier density noise amplitude is approximately given by ${\cal A}_{n} \propto T^2 F_0$,
where the quadratic temperature dependence is due to the  quantum capacitance. 
For $T \ll \Gamma/k_{\rm B}$, $F_0 \to \rho_{\rm T} k_{\rm B} T/( \pi \Gamma)$,
leading to  ${\cal A}_{n} \propto T^3$. With increasing temperature, $ T \gtrsim \Gamma/ k_{\rm B}$,
${\cal A}_{n}$ approaches ${\cal A}_{n} \propto T^2 F_0 = (\rho_{\rm T} /4) T^2$.
If the trap energy distribution center and the Fermi level are in the conduction band,
both amplitudes are related to correlators of  the second order in the trapped carriers density fluctuations, see Eqs. (\ref{eq:SIc_bdc}) and (\ref{eq:Sn_bdc}).
The case $\mu_0=\epsilon_{\rm T}= 5 \hbar v_{\rm D}/L$ is shown  in Figs.~\ref{fig:S_T}~(c) and (d).
In the considered temperature range, the critical current derivative with respect to the Fermi energy does 
not depend on the temperature and the parallel capacitance is dominated by the geometric capacitance, i.e. $C_\parallel \approx C_{\rm g}$, thus both
amplitudes follow the linear temperature dependence of $F_0$. 
Due to the Fabry-Perot oscillations of $I_{\rm c}$, the ratio 
$\mathcal{A}_{I_{\rm c}}/\mathcal{A}_{n} = (d I_{\rm c}/ d \mu_0)^2 / (2 C_{\rm g}^2/e^4)$
is to a certain extent tunable with the doping level. Moreover we note that 
both  ${\cal A}_{ I_{\rm c}}$  and ${\cal A}_{n}$ are considerably larger than at the CNP. 
For the carrier density noise the  scale factor is related to the capacitances ratio $(C_{\rm g}/C_{\rm Q})^2$, from Eqs. \eqref{eq:Sn_undoped} and \eqref{eq:Sn_bdc}.

{\bf Discussion}

Our analysis points out that short ballistic GJJs are sensitive  probes of microscopic
noise  underlying ballistic transport in 2D. 
In particular,  
we have shown that critical current noise probes either the second or higher order correlators of charge trapping 
centers fluctuations, by  tuning the doping level or the temperature. 
This result, obtained within a simple phenomenological model for discrete charge density fluctuations, 
highlights the GJJs potentialities to characterize non-Gaussian noise sources~\cite{paladino_rmp_2014,galperinPRL2006,paladinoPRL2002}.  
Independent measurements of critical current noise and carrier density noise could provide valuable insights on the underlying microscopic mechanisms and a cross-check of the McWorther's model applicability to GJJs.   
Charge carrier density noise may be probed via Hall voltage fluctuations measurements, an approach adopted in graphene  \cite{lu_prb_2014}.
Newly developed GJJ-based qubits may instead be employed as quantum sensors of critical current noise~\cite{pellegrino_mdpi_2019}.  
An important outcome of our analysis is the prediction of a linear $T$-dependence of the critical current noise amplitude
at sufficiently low-temperatures, independently of the details of the trap states energy distribution 
included in the density $\mathcal D_\Gamma(\epsilon)$. This behavior  arises in the regime $k_B T \ll {\min}[\Delta_0, \Gamma]$
from the factors $F_j$, defined by Eq.(\ref{eq:Fj}), when the critical current is approximately given by the zero doping value $1.33 e \Delta_0W/(\pi \hbar L)$. 
For characteristic values of GJJs on hBN, the fractional noise amplitude at the CNP
is approximately given by 
$ {\cal A}_{ I_{\rm c}} / I_{\rm c}^{2}  \sim  2 \pi  \times 10^{-7} \times  ( N_{\rm T}  \Delta/ \Gamma  ) \times (T/T_{\rm c})$, 
where $T_{\rm c}$ is the  critical temperature and  noise is measured in Hz$^{-1}$.
For finite doping instead the fractional amplitude is approximately one order of magnitude larger, 
$ {\cal A}_{ I_{\rm c}} / I_{\rm c}^{2}  \sim  2 \pi \times 10^{-6}   \times  ( N_{\rm T}  \Delta/ \Gamma  ) \times (T/T_{\rm c})$. 
An analogous temperature dependence observed in  Al/AlO$_x$/Al and Nb/AlO$_x$/Nb Josephson junctions,
scaling with the inverse junction area down to  $A_0 \approx 0.04~\mu {\rm m}^2$, has been attributed to ensembles of two-level fluctuators in the oxide barrier~\cite{NugrohoAPL2013,PottorfAPL2009}.
Superconducting qubits  are one of the forefront platforms for quantum state processing. In view of the 
relevance of hybrid superconducting circuits for quantum technologies, it interesting to benchmark critical current noise
in short ballistic-GJJ with figures in AlO$_x$-based Josephson junctions, where ${\cal A}_{I_{\rm c}}/I_{\rm c}^2 \approx 10^{-11} \times T/T_{\rm c}$ for junction's area $\sim A_0$\cite{NugrohoAPL2013}.
Assuming  a featureless $1/f$ spectrum due to $N_{\rm T}  \sim 10$ traps, 
a fractional noise amplitude comparable to the one in conventional Josephson junctions would imply 
a wide distribution of trap energies, $\Gamma \sim 10^{6} \times \Delta$ (for finite doping $\Gamma \sim 10^{5} \times \Delta$ ).
Within our phenomenological model, the  number of traps involved, $N_{\rm T}$, and the width of their energy distribution,
$\Gamma$,  are unknown parameters which could be estimated by fitting experimental data.


\vspace*{5mm}
\textbf{Methods}

In this work, we deal with the critical current noise of short and wide GJJs as a function of temperature and doping level.
In this regime, the supercurrent, defined in Eq.~(\ref{eq:superI}), can be expressed as~\cite{titov_prb_2006}
\begin{equation}
I(\phi)=-\frac{4 e W}{\hbar \pi}
\int_{0}^{\infty} d q \tanh\left[ \frac{\varepsilon(q,\phi)}{2 k_{\rm B} T} \right] \frac{\partial \varepsilon(q,\phi)}{\partial \phi}~,\nonumber
\end{equation}
where the summation over the transverse modes in Eq.~(\ref{eq:superI}) is replaced by the integration.
The integration above has been performed with Python numerical routines, 
in particular we have used the free and open-source library Scipy~\cite{scipy}. 
Similarly, to calculate the functions $F_j$s, defined in Eq.~(\ref{eq:Fj}) we have used the numerical integration routines included in Scipy.

\vspace*{5mm}


\bibliography{manuscript}
\vspace*{5mm}
\textbf{Acknowledgments.}
The authors thank G. G. N. Angilella, S. Kubatkin, M. Polini, F. Taddei, I. Torre, D. Vion for illuminating
discussions and fruitful comments on various stages of this work. 
This research was funded by
the project ``Linea di intervento 2'' of Dipartimento di Fisica e Astronomia ``Ettore Majorana'', Universit\`{a} di Catania.
\vspace*{5mm}

\newpage

\clearpage 
\setcounter{section}{0}
\setcounter{equation}{0}%
\setcounter{figure}{0}%
\setcounter{table}{0}%

\setcounter{page}{1}

\renewcommand{\thetable}{S\arabic{table}}
\renewcommand{\theequation}{S\arabic{equation}}
\renewcommand{\thefigure}{S\arabic{figure}}

 \makeatletter
\renewcommand\@bibitem[1]{\item\if@filesw \immediate\write\@auxout
    {\string\bibcite{#1}{S\the\value{\@listctr}}}\fi\ignorespaces}
\def\@biblabel#1{[S#1]}
 \makeatother

\begin{flushleft}
 \textbf{\Large Supplemental information for \\ ``1/f critical current noise in short ballistic graphene Josephson junctions''}
\end{flushleft}


\begin{center}
Francesco M.D. Pellegrino,$^{1\,,2}$
Giuseppe Falci,$^{1\,,2\,,3}$ and
Elisabetta Paladino,$^{1\,,2\,,3}$
\end{center}


\begin{flushleft}
{\small
$^1$\!{\it Dipartimento di Fisica e Astronomia ``Ettore Majorana'', \\Universit\`a di Catania, Via S. Sofia 64, I-95123 Catania,~Italy}

$^2$\!{\it INFN, Sez.~Catania, I-95123 Catania,~Italy}

$^3$\!{\it CNR-IMM, Via S. Sofia 64, I-95123 Catania,~Italy}
}
\end{flushleft}

\bigskip

\begin{figure}[h]
\centering
\includegraphics[width=.59\columnwidth]{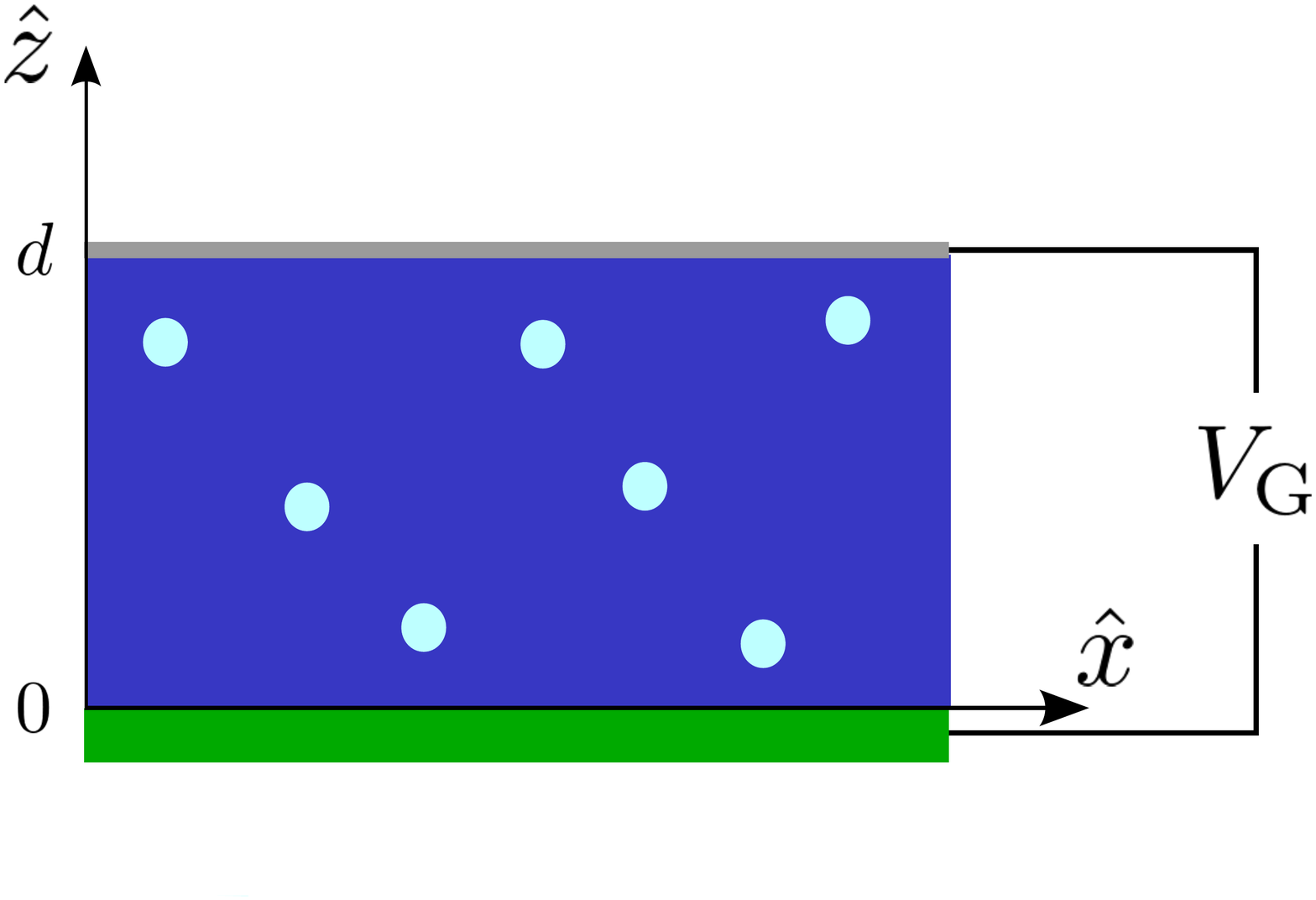}
\caption{Schematic of the device (side view). The device consists of a metal gate
(green), a substrate (blue), and a monolayer graphene (gray). 
Within the substrate, the cyan circles represent the electron traps. 
Between the monolayer graphene and the metal gate is forced a voltage drop $V_{\rm G}$ to tune charge carrier density in graphene.
}
\label{fig:side}
\end{figure}

As discussed in the main text, fluctuations of carrier density in the graphene insert of the ballistic GJJ 
are responsible for fluctuations of Andreev levels manifesting themselves as noise in the critical current.
Carrier density fluctuations are due to  charge trapping and release processes between graphene 
and carrier traps in the underlying substrate. This noise mechanism, typical of conventional semiconducting FET,
is commonly described by the McWorther  model~\cite{mcwhorther_1957}. We refer to the device sketched in Fig.~\ref{fig:side}. 
From bottom to top, it is composed by a metal gate (green), a substrate (blue), and a graphene monolayer (gray). In the  substrate, the cyan circles symbolize charge traps. 
In this supplemental material we derive the correlators of the charge trap populations  and 
evaluate the limiting forms of the  critical  current and charge carrier density power spectra for small and wide trap energy distributions and the asymptotic temperature behaviours.

Since the time between consecutive transitions is usually many times longer than the relaxation (equilibration) time 
of the crystal,  trapping and recombinations can be considered  as discrete Markov processes~\cite{kogan_buk}.
The occupancy number $X(i,t)$ of the trap labeled by the index $i$ at time $t$ is a random variable. 
Each trap can be empty ($X(i,t)=0$) or occupied by a single electron ($X(i,t)=1$), 
and it randomly switches between these states with time-independent rates (stationary process)~\cite{lax_rmp_1960}. 
The conditional probability that the trap $i$ at time $t$ has the occupation number $X(i,t)$  if 
the trap $j$ at time $t_0$ has the occupation number $X(j,t_0)$ is written as
\begin{equation}\label{eq:conditional}
 P[X(i,t)|X(j,t_0)]= \delta_{i, j}  P[X(i, t-t_0)|X(i, 0)]~,
\end{equation}
\begin{equation}
P[X(i, t)|X(i, 0)]= p_i[X(i, t)|X(i, 0)] + w_i (X(i, 0))~,
\end{equation}
where $w_i(X(i, 0))$ is the stationary probability of trap $i$ which depends on the  initial occupation $X(i, 0)$
as follows
$w_i(1)=f_{i}$ and $w_i(0)=1-f_{i}$. $p_i$ has the matrix form
\begin{equation}
 p_i[X(i, t)|X(i, 0)] =
\begin{bmatrix}
 p_i[0|0](t) &  p_i[0|1](t)  \\
 p_i[1|0](t) &  p_i[1|1](t)  \\
\end{bmatrix}~.
\end{equation}
The Kronecker delta appears in Eq.~(\ref{eq:conditional}) because different traps are uncorrelated.
The matrix $p_i$ is the solution of the  Kolmogorov equation~\cite{kogan_buk}
\begin{equation}
\frac{d p_i}{d t}=-M_{i} \cdot p_i~,
\end{equation}
where
\begin{equation}
M_{i} = 
\begin{bmatrix}
 \lambda_{i,00} & -\lambda_{i,11}\\
-\lambda_{i,00}  & \lambda_{i,11} 
\end{bmatrix}~,
\end{equation}
here $ \lambda_{i,00}$ and $\lambda_{i,11}$ are the transition rates for the processes $0 \to 1 $ and $1 \to 0$, respectively. 
Assuming equilibrium initial conditions
\begin{equation}
p_i[X_{i}( 0)|X_{i}( 0)] =
\begin{bmatrix}
f_i & -(1-f_i) \\
- f_i   & 1-f_i 
\end{bmatrix}~,
\end{equation}
the solution of the Kolmogorov equation is given by
\begin{equation}\label{eq:p}
p_i[X_{i}( 0)|X_{i}( 0)] =
\begin{bmatrix}
f_i & -(1-f_i) \\
- f_i   & 1-f_i  
\end{bmatrix} e^{-\gamma_i t}~,
\end{equation}
where $\gamma_i=\lambda_{i,00} +\lambda_{i,11} $ is the overall switching rate between the two states
of the stochastic process. In the following we'll specify the form of $\gamma_i$.
Due to Markovianity, the multi-time correlators reduce to two-points correlation function
\begin{equation}\label{eq:condchain}
 P[X(i_N, t_N)|X(i_{N-1}, t_{N-1});\ldots ;X(i_0, t_0)]=  P[X(i_N, t_N)|X(i_{N-1}, t_{N-1})]~.
\end{equation}
The density of populated traps per unit volume and energy, ${\cal N}_{\rm T}(\epsilon,\bm R,t)$,
fluctuates around its average value ${\cal N}_{{\rm T}0}$ and  can be expressed as 
\begin{equation}
 {\cal N}_{\rm T}(\epsilon,\bm R,t)\equiv{\cal N}_{{\rm T}0}(\epsilon,\bm R)+\delta {\cal N}_{\rm T}(\epsilon,{\bm R},t )~.
\end{equation}
Assuming that trap $i$ is located at position $\bR_i$ and that the energy of the occupied trap is $\epsilon_i$ (evaluated
with respect to the CNP), the average value can be expressed as
\begin{equation} 
 {\cal N}_{{\rm T}0}(\epsilon,\bm R)\equiv\sum_{i=1}^{M_{\rm T}} \delta(\epsilon-\epsilon_i) \delta(\bR-\bR_i) f_i~,
\end{equation}
and the fluctuations as
\begin{equation}\label{eq:nt_random}
\delta {\cal N}_{\rm T}(\epsilon,\bR,t )\equiv \sum_{i=1}^{M_{\rm T}} \delta(\epsilon-\epsilon_i) \delta(\bR-\bR_i) [X(i,t)- f_i]~,
\end{equation}
where $M_{\rm T}$ is the total amount of traps. 
From here on, we assume that the stationary probability coincides with the equilibrium occupation function 
\begin{equation}
 f_i=f(\epsilon_i,\bR_i)=f_{\rm D}(\epsilon_i-\mu_0)~,
\end{equation}
where $f_{\rm D}(x)=1/\{1+\exp{[x/(k_{\rm B} T)]}\}$ is the Fermi-Dirac distribution function and $\mu_0$ is the Fermi level.
Under these conditions it is easy to evaluate the  average density of populated traps per unit volume and energy and the multi-time correlators which read 
\begin{equation}\label{eq:n}
 \langle \delta  {\cal N}_{\rm T}(\epsilon,\bR,t ) \rangle = 0~,
\end{equation}
\begin{align}\label{eq:nn}
& \langle \delta {\cal N}_{\rm T}(\epsilon_1,\bR_1,t_1 )\delta {\cal N}_{\rm T}(\epsilon_0,\bR_0,t_0 ) \rangle = 
\\
& \delta(\bR_1 -\bR_0) \delta(\epsilon_1-\epsilon_0)  {\cal D}(\epsilon_0,\bR_0) 
f_{\rm D}(\epsilon_0- \mu_0)[1-f_{\rm D}(\epsilon_0- \mu_0)]\exp[-\gamma(\epsilon_0, z_0)(t_1-t_0)]
~,\nonumber
\end{align}
\begin{align}\label{eq:nnn}
 &\langle \delta {\cal N}_{\rm T}(\epsilon_2,\bR_2,t_2 ) \delta {\cal N}_{\rm T}(\epsilon_1,\bR_1,t_1 ) \delta {\cal N}_{\rm T}(\epsilon_0,\bR_0,t_0 ) \rangle = 
\delta(\bR_2-\bR_1)\delta(\bR_1 -\bR_0)  \\
 &\times \delta(\epsilon_2-\epsilon_1)\delta(\epsilon_1-\epsilon_0) 
 {\cal D}(\epsilon_0,\bR_0) ) f_{\rm D}(\epsilon_0- \mu_0)
[1-f_{\rm D}(\epsilon_0- \mu_0)]  [1-2 f_{\rm D}(\epsilon_0- \mu_0)]  \nonumber \\
 &\times \exp[-\gamma(\epsilon_0,z_0)(t_2-t_0)]~, \nonumber
\end{align}
\begin{align}\label{eq:nnnn}
& \langle \delta {\cal N}_{\rm T}(\epsilon_3,\bR_3,t_3 ) \delta {\cal N}_{\rm T}(\epsilon_2,\bR_2,t_2 ) \delta {\cal N}_{\rm T}(\epsilon_1,\bR_1,t_1 ) 
\delta {\cal N}_{\rm T}(\epsilon_0,\bR_0,t_0 )  \rangle = \\
&\langle \delta {\cal N}_{\rm T}(\epsilon_3,\bR_3,t_3 ) \delta {\cal N}_{\rm T}(\epsilon_2,\bR_2,t_2 ) \rangle 
\langle \delta {\cal N}_{\rm T}(\epsilon_1,\bR_1,t_1 )\delta {\cal N}_{\rm T}(\epsilon_0,\bR_0,t_0 )  \rangle \nonumber\\
&+ \delta(\bR_3 -\bR_2)\delta(\bR_2 -\bR_1)\delta(\bR_1 -\bR_0) \delta(\epsilon_3-\epsilon_2) \delta(\epsilon_2-\epsilon_1)
\delta(\epsilon_1-\epsilon_0) \nonumber \\ 
&\times {\cal D}(\epsilon_0,\bR_0)   f_{\rm D}(\epsilon_0- \mu_0)
[1-f_{\rm D}(\epsilon_0- \mu_0)]  [1-2 f_{\rm D}(\epsilon_0- \mu_0)]^2 
\exp[-\gamma(\epsilon_0,z_0)(t_3-t_0)]~,\nonumber
 \end{align}
where  we have explicitly indicated that the switching rates $\gamma_i$ may  depend on the energy $\epsilon_i$ and 
on the distance of the trap from the graphene layer, expressed by the $z$ coordinate.
Moreover $ {\cal D} ( \epsilon, {\bm R}) $ represents the number of trap states per unit volume and energy 
at position ${\bm R}$ for energy $\epsilon$
\begin{equation}
 {\cal D} ( \epsilon, {\bm R})  \equiv \sum_{i=1}^{M_{\rm T}} \delta(\epsilon-\epsilon_i) \delta(\bR-\bR_i)~.
\end{equation}
We assume  that  traps are randomly distributed in the substrate and we express the deviations of the voltage 
drop $V_{\rm T}$, defined in Equation (8) of the main text, from its equilibrium value as
\begin{equation}\label{eq:dVt}
\delta V_{\rm T} (t) = \frac{4 \pi e}{\epsilon_{\rm r}} \int \frac{d {\bm r} }{L W}\int_0^d dz z \int_{-\Lambda}^{\Lambda} \delta {\cal N}_{\rm T}(\epsilon,\bR,t )~, 
\end{equation}
where $\bR=(\br,z)$.
By using the correlators given in Eqs.~(\ref{eq:n})-(\ref{eq:nnnn}) and  Eq.~(\ref{eq:dVt}), it is
easy to show that the correlators of the voltage drop due to charge traps population fluctuations can be written as
\begin{equation}\label{eq:VtVt}
 \langle \delta V_{\rm T}(t)  \delta V_{\rm T} (0) \rangle= 
 \frac{e^2  }{C_{\rm g}^2 L W} G_0(t)~,
\end{equation}
 \begin{equation}\label{eq:Vt2Vt}
 \langle [\delta V_{\rm T}(t)]^2  \delta V_{\rm T} (0) \rangle=  \langle \delta V_{\rm T}(t)  [\delta V_{\rm T} (0)]^2 \rangle=\frac{e^3}{C_{\rm g}^3 L^2 W^2}  G_1(t)~,
 \end{equation}
\begin{equation}\label{eq:Vt2Vt2}
 \langle [\delta V_{\rm T}(t)]^2  [\delta V_{\rm T}(0)]^2 \rangle -  \langle [\delta V_{\rm T}(t)]^2\rangle \langle  [\delta V_{\rm T}(0)]^2 \rangle=
\frac{e^4 }{C_{\rm g}^4 L^3 W^3} G_2(t)~,
\end{equation}
where $C_{\rm g}=\epsilon_{\rm r}/(4 \pi d)$ is the geometric capacitance and the functions $G_j(t)$
read
\begin{equation}
 G_j (t)\equiv \int_{-\Lambda}^\Lambda d \epsilon {\cal D}(\epsilon) 
  f_{\rm D}( \epsilon -\mu_0)[1-f_{\rm D}( \epsilon -\mu_0)][1-2f_{\rm D}( \epsilon -\mu_0)]^j
g_{j+2}(\epsilon,t)~,
\end{equation} 
\begin{equation}\label{eq:gn}
  g_j(\epsilon,t) \equiv \int_0^d d z \left(\frac{z}{d}\right)^j\exp[- \gamma(\epsilon,z) t ]~,
\end{equation}
having assumed that the density of trap states does not depend on $\bR$ and indicated it as
 ${\cal D}(\epsilon)$.
The switching rate depends on the trap distance from the graphene layer as ~\cite{mcwhorther_1957,balandin_nn_2013}
\begin{equation}\label{eq:tunnel}
 \gamma(\epsilon,z)=\gamma_0 \exp[-|z-d|/\ell(\epsilon)]+\gamma_0^\prime \exp[-|z|/\ell^\prime(\epsilon)]~,
\end{equation}
where we distinguish tunneling processes related to the graphene channel, characterized by $\gamma_0$ and $\ell(\epsilon)$, 
and tunneling processes related to the gate channel, characterized by $\gamma_0^\prime$ and $\ell^\prime(\epsilon)$. 
The lengths $\ell(\epsilon)$ and $\ell^\prime(\epsilon)$ can depend on the trap energy   but are such that $\ell(\epsilon), \ell^\prime(\epsilon) \ll d$.
Under these conditions, the Fourier cosine transforms of the correlators, for frequencies $\omega/( 2 \pi)$  within the range
$\gamma_0 e^{- d/\ell(\epsilon)}$, $\gamma_0^\prime e^{- d/\ell^\prime(\epsilon)} \ll \omega \ll \gamma_0,\gamma_0^\prime$, 
are given by
\begin{equation}\label{eq:gnomega}
 \int_0^\infty \frac{d t}{\pi} \cos(\omega t)  g_j(\epsilon, t) \approx \frac{\ell(\epsilon)}{2 \omega}~.
\end{equation}
Therefore  we can write
\begin{equation}\label{eq:BigG}
 \int_0^\infty \frac{d t}{\pi} \cos(\omega t)  G_j (t) \approx  \frac{\ell_0  F_j  }{2 \omega}~,
\end{equation}
 where $\ell_0=\ell(\mu_0)$ and 
%
\begin{equation}
 F_j \equiv  \int_{-\Lambda}^\Lambda d \epsilon  {\cal D}(\epsilon) \frac{\ell(\epsilon)}{\ell_0} 
 f_{\rm D}( \epsilon -\mu_0)[1-f_{\rm D}( \epsilon -\mu_0)][1-2f_{\rm D}( \epsilon -\mu_0)]^j~.
\end{equation}
The term ${\cal D}(\epsilon)  \ell(\epsilon)/\ell_0 $ represents a renormalized density of trap states.
In the main text, we neglect the energy dependence of $\ell(\epsilon)$, i.e.
\begin{equation}
 F_j =  \int_{-\Lambda}^\Lambda d \epsilon  {\cal D}(\epsilon) 
 f_{\rm D}( \epsilon -\mu_0)[1-f_{\rm D}( \epsilon -\mu_0)][1-2f_{\rm D}( \epsilon -\mu_0)]^j~.
\end{equation}
By exploiting Eq.~(\ref{eq:BigG}), one can write  the following compact form for the Fourier cosine transforms of
$ \delta V_{\rm T}(t) $ correlation functions
\begin{equation}\label{eq:VtVtomega}
 \int_0^\infty \frac{d t}{\pi} \cos(\omega t) \langle \delta V_{\rm T}(t)  \delta V_{\rm T} (0) \rangle\approx 
 \frac{e^2  }{C_{\rm g}^2 L W} \frac{ \ell_0 F_0 }{2 \omega} 
~,
\end{equation}
\begin{equation}\label{eq:Vt2Vtomega}
 \int_0^\infty \frac{d t}{\pi} \cos(\omega t)  \langle [\delta V_{\rm T}(t)]^2  \delta V_{\rm T} (0) \rangle  \approx
\frac{e^3 }{C_{\rm g}^3 L^2 W^2}  
\frac{ \ell_0 F_1 }{2 \omega} 
~,
 \end{equation}
\begin{equation}\label{eq:Vt2Vt2omega}
 \int_0^\infty \frac{d t}{\pi} \cos(\omega t)  \left\{  \langle [\delta V_{\rm T}(t)]^2  [\delta V_{\rm T}(0)]^2 \rangle -  \langle [\delta V_{\rm T}(t)]^2\rangle \langle  [\delta V_{\rm T}(0)]^2 \rangle  \right\} 
 \approx
\frac{e^4 }{C_{\rm g}^4 L^3 W^3} \frac{\ell_0 F_2 }{2 \omega}~.
 \end{equation}
In the main text we related fluctuations of the critical current and of the carrier density to correlators of the 
Fermi level of different orders, see Eqs.~(4) and (22). Since   
there is a linear relation between the deviation of the Fermi level from its equilibrium value
and  $\delta V_{\rm T}$, expressed in Eq.~(10) in the main text, it is easy to show that
\begin{equation}\label{eq:deltaIc}
 \delta I_{\rm c}(t) \approx   -\frac{d I_{\rm c}(\mu_0)}{d \mu_0}  \frac{ e C_{\rm g}}{C_{\rm g}+C_{\rm Q}} \delta V_{\rm T}(t)+  
 \frac{d^2 I_{\rm c}(\mu_0)}{d \mu_0^2} \left( \frac{ e C_{\rm g}}{C_{\rm g}+C_{\rm Q}} \right)^2  \frac{[\delta  V_{\rm T}(t)]^2}{2}~,
\end{equation}
and
\begin{equation}\label{eq:deltan}
\delta n (t)=-\frac{C_{\rm Q} C_{\rm g}}{e (C_{\rm g}+C_{\rm Q})}  \delta V_{\rm T}(t) +  \frac{d C_{\rm Q}}{d \mu_0} \left( \frac{  C_{\rm g}}{C_{\rm g}+C_{\rm Q}} \right)^2  \frac{[\delta  V_{\rm T}(t)]^2}{2}~,
\end{equation}
where $C_{\rm Q}=e^2 (d n_0/d \mu_0)$ is the quantum capacitance.
By using Eqs.~(\ref{eq:VtVtomega})-(\ref{eq:deltan}) one finds the power spectra of the critical current and of the charge carrier density in the following analytical forms
\begin{align}
{\cal S}_{ I_{\rm c}} (\omega)&\equiv \int_0^\infty \cos(\omega t) \langle \delta  I_{\rm c}(t) \delta  I_{\rm c}(0) \rangle \\
& \approx  \Bigg[
  \Bigg(  \frac{d I_{\rm c}}{d \mu_0}  \Bigg)^2 F_0 
 - \Bigg(  \frac{d I_{\rm c}}{d \mu_0}  \Bigg) \Bigg(  \frac{d^2 I_{\rm c}}{d \mu_0^2}  \Bigg) \varepsilon_{\rm Q} F_1  
 +\Bigg(  \frac{d^2 I_{\rm c}}{d \mu_0^2}  \Bigg)^2   \frac{\varepsilon_{\rm Q}^2}{4}  F_2
\Bigg]
\varepsilon_{\rm Q}^2 \frac{L W \ell_0 }{2 \omega}~,\nonumber \\
{\cal S}_{n} (\omega)&\equiv \int_0^\infty \cos(\omega t) \langle \delta n(t) \delta  n(0) \rangle  \\
&\approx
\Bigg[
C_{\rm Q}^2 F_0 
 - C_{\rm Q} \frac{d C_{\rm Q}}{ d \mu_0}  \varepsilon_{\rm Q} F_1  
 +\Bigg(  \frac{d C_{\rm Q}}{ d \mu_0}  \Bigg)^2   \frac{\varepsilon_{\rm Q}^2}{4}  F_2
\Bigg]
\varepsilon_{\rm Q}^2 \frac{L W \ell_0 }{2 e^4 \omega}~, \nonumber
\end{align}
where $\varepsilon_{\rm Q}= e^2/(C_\parallel L W)$, and $C_\parallel=C_{\rm g}+ C_{\rm Q}$.
In both power spectra ${\cal S}_{ I_{\rm c}}$ and ${\cal S}_{n}$, all microscopic information concerning the traps are 
included in the $F_j$ functions.

In the main text, we choose a density of trap states with the quite general Lorentzian form
${\cal D}_\Gamma(\epsilon)=\pi^{-1}\rho_{\rm T} \Gamma/[(\epsilon-\epsilon_{\rm T})^2 +\Gamma^2]$. 
The functions $F_j$ can be evaluated in analytic form in the limiting regimes of temperatures much smaller or
larger than the width of the trap energy distribution. Here we derive these asymptotic behaviors.  
To this aim we perform a suitable change of variable and we rewrite the $F_j$ functions as
\begin{equation}
\label{change_v}
 F_j = \frac{\rho_{\rm T}}{\pi}  \int_{-\infty}^\infty d \xi  \frac{\Gamma/(k_{\rm B} T)}{\xi^2+\Gamma^2/(k_{\rm B} T)^2}
\frac{\tanh^{j}(\xi + \frac{\epsilon_{\rm T} -\mu_0}{2k_{\rm B} T})}{4 \cosh^2(\xi + \frac{\epsilon_{\rm T} -\mu_0}{2k_{\rm B} T})}~,
\end{equation}
having assumed a cut-off energy $\Lambda$  much larger than any other energy scale.
From Eq. (\ref{change_v}) it is easy to see that in the low temperature regime $\Gamma/ k_{\rm B}T \gg 1$, 
one has 
\begin{equation}
 F_j \to \frac{\rho_{\rm T}}{\Gamma \pi}  \int_{-\infty}^\infty d \xi  
\frac{\tanh^{j}(\xi + \frac{\epsilon_{\rm T} -\mu_0}{2k_{\rm B} T})}{4 \cosh^2(\xi + \frac{\epsilon_{\rm T} -\mu_0}{2k_{\rm B} T})}~,
\end{equation}
which leads to a linear temperature dependence for the even correlators and vanishing of the odd correlators
\begin{subequations}\label{eq:FnD0}
\begin{eqnarray}
  F_0 &\to&  \frac{\rho_{\rm T}}{\pi \Gamma} k_{\rm B} T ~, \\
  F_1 &\to& 0~,\\
  F_2 &\to& \frac{\rho_{\rm T}}{3 \pi \Gamma} k_{\rm B} T~.
\end{eqnarray}
\end{subequations} 
In the high temperature regime $\Gamma/ k_{\rm B}T \ll 1$ from Eq. (\ref{change_v}) we obtain 
\begin{equation}\label{eq:Fndelta}
 F_j \to \rho_{\rm T} \int_{-\infty}^\infty d \xi  \delta(\xi)
\frac{\tanh^{j}(\xi + \frac{\epsilon_{\rm T} -\mu_0}{2k_{\rm B} T})}{4 \cosh^2(\xi + \frac{\epsilon_{\rm T} -\mu_0}{2k_{\rm B} T})}=
\rho_{\rm T} \frac{\tanh^{j}(\frac{\epsilon_0-\mu_0}{2k_{\rm B} T})}{4 \cosh^2(\frac{\epsilon_{\rm T}-\mu_0}{2k_{\rm B} T})}~.
\end{equation}
These dependencies are at the origin of the low and high temperature forms of the critical current and charge carrier 
density noise amplitudes discussed in the main text after Figure 5.

\end{document}